\documentclass[twocolumn,aps,prc]{revtex4}

\usepackage{float}
\usepackage{url}
\usepackage{bm}

\usepackage[utf8]{inputenc}
\usepackage{amsfonts}
\usepackage{amsmath}
\usepackage{amssymb}
\usepackage{epsfig}
\usepackage{bm}

\usepackage{fixme}
\usepackage{color}
\usepackage{cancel}

\graphicspath{{./figures/}}

\begin{document}
\title{Three-body structures of low-lying nuclear states of $^8$Li}

\author{E. Garrido$^1$, A.S. Jensen$^2$}
\affiliation{$^1$Instituto de Estructura de la Materia, CSIC, Serrano 123, E-28006 Madrid, Spain}

\affiliation{$^{2}$Department of Physics and Astronomy, Aarhus University, DK-8000 Aarhus C, Denmark}

\date{\today}

\begin{abstract}
The four nucleons in $^8$Li outside the $\alpha$-particle
($\alpha=^4$He) can be divided into pairs of one neutron ($n$) and 3
nucleons in the triton ($t=^3$H), or 2 in the deuteron ($d=^2$H) and
two neutrons in a dineutron ($^2n$).  The corresponding three-body
structure, $\alpha$+$t$+$n$ or $\alpha$+$d$+$^2n$, are suggested to
describe the bulk part of the low-energy ($<10$~MeV) states of $^8$Li.
Several breakup thresholds influence the structures and possible
decays.  We calculate the three-body structures of the various
$J^{\pi}$ states, where different clustering appear, e.g. $^7$Li*+$n$,
$^6$Li*$+^2n$, $^6$He*$+d$.  The experimental $^8$Li spectrum can be
reproduced with fine tuning by a three-body potential parameter.
Three unobserved $0^+$ and an excited 2$^+$ states are found.  All states appear as bound
states or resonances.  The lowest or highest energies have cluster
structures, $\alpha$+$t$+$n$ or $\alpha$+$d$+$^2n$, respectively.  We
give calculated energy and width (if possible), geometry, and partial
wave decomposition for all states.
\end{abstract}

\maketitle

\section{Introduction}

The structure of a nucleus, a system  consisting of neutrons and protons, is in
principle a many-body problem \cite{Bethe1956,Schmid1962}.  For many light nuclei, the
no-core shell model has been applied rather successfully to describe
the low-lying states \cite{Navratil2009}. Also Quantum and Green's function
Monte Carlo methods have been applied on nuclear states of $^8$Li
\cite{Wiringa2000}.  However, beside the intrinsic many-body challenges,
especially two problems are not easy to overcome, that is,  the
strong tendency for nucleons to form clusters, and  to account
for the continuum. First, the competition between cluster
configurations and nucleonic motion requires a very large
single-particle basis, leading in consequence to extensive
calculations.  Second, without inclusion of the
continuum, it is not possible to describe the decay mechanism and the
widths of resonances.

Both problems disappear in few-body calculations \cite{Zhukov1993,Fynbo2009}, where inert and
effectively point-like bound clusters are assumed from the beginning.
The numerical treatment of all relative cluster motion is by definition
accurate and complete.  The numerical difficulties, as well as
interest, increase when breakup thresholds appear in the spectrum.  The choice of clusters is
crucial, since a state in the vicinity of a threshold corresponding to
a given cluster structure, tends to have strong similarity to this
structure. Thus, different thresholds imply different cluster
structures, and in turn therefore non-identical cluster configurations
must contribute.

The light nucleus, $^{8}$Li, is a promising candidate with all theses
features \cite{Tilley2004,Baye1994}.  Its structure is from the
beginning a nuclear eight-body problem.  However, four of the
particles have a great desire to combine into the very stable
$\alpha$-particle, leaving a five-body problem.  The four particles, 3
neutrons and one proton, outside the $\alpha$-particle can combine in
different ways.  Due to the large neutron separation energy of the
triton, 6.26~MeV \cite{Purcell2010}, it is reasonable to think that
the preferred structure of the lowest states of $^{8}$Li is a
three-body configuration made of $\alpha$, $t$ and $n$.  Here $\alpha$
and $t$ could combine to form $^{7}$Li, which with the neutron form a
two-body system. The neutron can also be equally shared or attached to
either $\alpha$ or $t$, resembling $^{5}$He or $^{4}$H.  All such
states would be contained in the same full three-body calculation with
$\alpha+t+n$.

Other three-body configurations, allowed at higher energies, are also
possible. In particular, we can think of $^{6}$He($\alpha+n+n$)+$n+p$
and $^{6}$Li($\alpha+d$)+$n$+$n$.  These two configurations can be
combined in three-body structures, $\alpha$+$d$+$^2n$, where the
$\alpha$-particle can combine with either the dineutron, $^2n$, to
form $^6$He, or with the deuteron, $d$, to form $^6$Li.  This
three-body configuration has the advantage to permit $^8$Li
states described as built of various combinations of excited states of
$^6$Li or $^6$He.  This is not possible in a $^6$Li+$n$+$n$ or
$^6$He+$n$+$p$ three-body structure, where only one structureless
state of $^6$Li or $^6$He are allowed.  Such three-body calculations
are an efficient alternative, and tremendously much easier, than
solving the full five-body problem.

The purpose of this work is to investigate how well these two
three-body configurations, $\alpha$+$t$+$n$ and $\alpha$+$d$+$^2n$,
can describe the $^8$Li low-energy spectrum. We first emphasize that
these two structures automatically restrict the investigations to
$^8$Li states with total isospin 1.  The $\alpha$-particle has zero
isospin, the isospin $\frac{1}{2}$ of the triton and the isospin
$\frac{1}{2}$ of the neutron can produce only total isospin 0 or 1,
and 0 is not possible because $^8$Li has a different number of
neutrons and protons.  For the $\alpha$+$d$+$^2n$ configuration, both
$\alpha$ and $d$, have zero isospin, and the total isospin is given by
the isospin $1$ of the dineutron.  A total isospin $2$ for $^8$Li is
also possible, but that would require a structure like
$\alpha$+$np$+$^2n$, with the $np$ system in an isospin 1 state.  This
type of structure is not considered here and left for future work.

The paper is structured as follows. In Section~\ref{sec:2bd} we give the 
two-body interactions used in the three-body calculations. The
three-body method is briefly summarized in Section~\ref{sec3}.
Sections~\ref{sec:atn} and \ref{sec:ad2n} are devoted to the results obtained for the $^8$Li
states with each of the two considered three-body configurations,
where the structure of each of the computed states is analyzed.
A comparison with previous calculations is discussed in Section~\ref{sec:comp}, 
and
Section~\ref{sec:sum} contains summary and conclusions.

\section{Two-body interactions}
\label{sec:2bd}
The description of $^8$Li as a three-body system,
e.g. $\alpha$+$t$+$n$ or $\alpha$+$d$+$^2n$, requires specification of
the three different two-body interactions involved in each of the
cases.  The radial dependence is assumed to be sums of Gaussians,
$V(r)=Se^{-r^2/b^2}$, depending on spin and angular momentum of the
interactions.  All the state-dependent strengths, $S$, and ranges,
$b$, will be determined to reproduce the available experimental
information for each of the two-body subsystems.  The details of these
interactions for the two three-body configurations considered are
given in the following subsections.

\subsection{$\alpha$+$t$+$n$ configuration}
\label{sec:2bda}

\paragraph{The triton-neutron potential:} The  $t$-$n$ interaction is taken from  Ref.\cite{Diego2007}, where it is constructed to describe the unbound 
nuclei, $^4$H and $^5$H. The interaction has the form:
\begin{equation}~
V_{tn}^{(\ell)}(r)=V_c^{(\ell)}(r)+V_{js}^{(\ell)}(r)\bm{j}_n\cdot \bm{s}_t+V_\mathrm{so}^{(\ell)}(r) \bm{\ell}\cdot\bm{s}_n,
\label{pot2ba}
\end{equation}
where $\bm{\ell}$ is the relative orbital angular momentum between the triton and the neutron, $\bm{s}_t$ and $\bm{s}_n$ are their spins, 
respectively, and $\bm{j}_n=\bm{\ell}+\bm{s}_n$. The total angular momentum of the $t$-$n$ two-body system comes from the
coupling of $\bm{j}_n$ and $\bm{s}_t$.

As shown in Ref.~{\cite{Garrido2003}}, the choice of the operators
made in Eq.(\ref{pot2ba}) is such that $\bm{j}_n$ is the conserved
(for given $\bm{\ell}$) total angular momentum of a particle
surrounding a core.  This allows obeying the Pauli principle, since
$\bm{j}_n$ then characterizes nucleon-shells in the core, where only
unoccupied core-states are available for the valence nucleons.

\begin{table}[t!]
  \begin{tabular}{|cccc|ccc|cc|}
\hline    
 \multicolumn{1}{|c}{}
 & \multicolumn{8}{c|}{triton-neutron interaction} \\ \hline
 \multicolumn{1}{|c}{}  &
\multicolumn{3}{c|}{$V_c^{(\ell)}$ } &
\multicolumn{3}{c}{$V_{js}^{(\ell)}$ } &
\multicolumn{2}{|c|}{$V_\mathrm{so}^{(\ell)}$ } \\ \hline
$\ell\,\,\,\,$ & $0$ & $1$ & $2$ & $0$ & $1$ & $2$ & $1$ & $2$ \\
\hline
$S_1\,\,\,\,$  & 20.17 & $-400.01$ & 10.04 & $-1.45$ & $-237.77$ & 2.80 & $ -135.98$ & $-2.15 $\\
$b_1\,\,\,\,$  & 3.59 & 2.62   & 3.62 & 7.39 & 1.44   & 3.38 &  2.09   & 3.68 \\
$S_2\,\,\,\,$  & --   & 400.81 & --   & --   & 108.87 & --   &  197.56 & -- \\
$b_2\,\,\,\,$  & --   & 2.55   & --   & --   & 1.70   & --   &  1.82   & -- \\
\hline
\end{tabular}
\caption{For the triton-neutron interaction, strengths $S_i$ and ranges $b_i$ of the two Gaussians ($i=1,2$) for the central, $V_c^{(\ell)}$, $js$-term, $V_{js}^{(\ell)}$, and
spin-orbit, $V_\mathrm{so}^{(\ell)}$, potentials in Eq.(\ref{pot2ba}). The strengths are given in MeV and the ranges in fm.}
\label{tab1}
\end{table}

The central, $V_c^{(\ell)}$, spin-spin, $V_{js}^{(\ell)}$, and
spin-orbit, $V_\mathrm{so}^{(\ell)}$, functions are chosen to be
$\ell$-dependent and given as a sum of two Gaussians.  The strengths
and ranges of the Gaussians have been obtained to reproduce, together
with the Coulomb potential, the experimental $s$-, $p$-, and $d$-waves
of the triton-proton phase shifts (see \cite{Diego2007}
for details).  The neutron-triton interaction is therefore extracted
simply by switching off the Coulomb repulsion.

In Table~\ref{tab1} we give the strengths and ranges of the Gaussians describing $V_c^{(\ell)}$,  $V_{js}^{(\ell)}$, and $V_\mathrm{so}^{(\ell)}$ for $\ell=$0, 1, and 2. Note that the
$s$-wave potential is repulsive, which prevents the neutron from binding into the $s_{1/2}$ shell, which is fully occupied by the two neutrons in the triton, and therefore Pauli forbidden for the third neutron.

The potentials given in Table~\ref{tab1} give rise to four $p$-wave
resonances, the pairs $\{0^-, 1^-\}$ and $\{1^-, 2^-\}$,
respectively coming from the coupling of a $p_{1/2}$ and a $p_{3/2}$
neutron to the spin $\frac{1}{2}$ of $t$. The computed energies, $E_R$, and
widths, $\Gamma_R$, of these four resonant states (obtained as poles
of the $S$-matrix) are $(E_R,\Gamma_R)$= (0.77, 6.72) MeV, (1.15,
6.38) MeV, (1.15, 3.49) MeV, and (1.22, 3.34) MeV, respectively. These
values agree with the ones obtained in other calculations of the
resonances as poles of the $S$-matrix \cite{Arai2003}, but they are
systematically smaller than the values obtained using $R$-matrix
analysis \cite{Tilley1992} (see Ref.\cite{Diego2007} for a
discussion about this issue.)

\begin{table}[t!]
\begin{tabular}{|cccc|cc|}
\hline
  & \multicolumn{5}{c|}{$\alpha$-neutron interaction} \\
\hline
   &
\multicolumn{3}{c|}{$V_c^{(\ell)}$ } &
\multicolumn{2}{c|}{$V_\mathrm{so}^{(\ell)}$ } \\
\hline
$\ell\,\,\,\,$ & $0$ & $1$ & $2$  & $1$ & $2$ \\
\hline
$S\,\,\,\,$  & 48.0 & $-47.40$ & $-21.93$ &   $-25.49$ & $-25.49$ \\
$b\,\,\,\,$  & 2.33 & 2.30   & 2.03 &   1.72   & 1.72 \\ \hline
\multicolumn{6}{c}{ } \\ \hline
& \multicolumn{5}{c|}{$\alpha$-triton interaction} \\ 
\hline
   &
\multicolumn{3}{c|}{$V_c^{(\ell)}$ } &
\multicolumn{2}{c|}{$V_\mathrm{so}^{(\ell)}$ } \\
\hline
$\ell\,\,\,\,$ & $0$ & $1$   & $3$  & $1$ & $3$\\
\hline
$S\,\,\,\, $  & 30.0 & $-17.97$ &  $-31.47$, $-57.99$ &  $-0.72$  & $20.98$, $-29.0$\\
$b\,\,\,\,$  & 3.0 & 4.0   &   2.55, 2.32 & 4.0   &  2.55, 2.32\\
\hline
\end{tabular}
\caption{Strength ,$S$, and range, $b$, of the Gaussian central, $V_c^{(\ell)}$, and
spin-orbit, $V_\mathrm{so}^{(\ell)}$, potentials in Eq.(\ref{pot2bb}) for the $\alpha$-neutron (upper part) and
$\alpha$-triton  (lower part) systems. For $\ell=3$ in the $\alpha$-triton case, the potential is given as the sum of two Gaussians.
The strengths are given in MeV and the ranges in fm.}
\label{tab2}
\end{table}

\paragraph{The $\alpha$-neutron potential:}  The $\alpha$-neutron interaction is taken as in Eq.(\ref{pot2ba}), which, due to zero spin of the $\alpha$-particle, reduces to
\begin{equation}
V_{\alpha n}^{(\ell)}(r)=V_c^{(\ell)}(r)+V_\mathrm{so}^{(\ell)}(r)\bm{\ell}´\cdot\bm{s}_n,
\label{pot2bb}
\end{equation}
where the $\ell$-dependent central and spin-orbit functions again are considered with Gaussian shape. 
The strength, $S$, and the range, $b$, of the Gaussians are adjusted separately for $s$, $p$, and $d$-waves, and take the values employed in Ref.~\cite{Cobis1998} and given in the upper part of Table~\ref{tab2}. As in the triton-neutron case, the repulsive $s$-wave 
potential guarantees fulfillment of the Pauli principle, which forbids the neutron to bind into the $s_{1/2}$ shell, which is fully occupied in the $\alpha$-particle. The equivalence 
between this sort of shallow potentials and deep potentials in principle allowing the binding into a Pauli forbidden state is investigated in  Ref.~\cite{Garrido1999}.

These potentials reproduce the experimental low-energy ($E\lesssim 20$
MeV) $s$-, $p$-, and $d$-wave $\alpha$-neutron phase shifts, as well
as the experimental energy and width of the $\frac{3}{2}^-$ and $\frac{1}{2}^-$
resonances in $^5$He. At the three-body level, these Gaussian
potentials have been successfully employed to describe the main
properties of the $^6$He ground state \cite{Cobis1998}. On top of
this, these potentials have also been used to describe the momentum
distributions of the fragments after breakup of a $^6$He projectile
\cite{Garrido1998c} or the $2^+$ resonance in $^6$He
\cite{Fedorov2003}.

\paragraph{The $\alpha$-triton potential:}
The interaction between the $\alpha$-particle and triton is also considered to contain  $\ell$-dependent central and spin-orbit terms, as given in Eq.(\ref{pot2bb}), where the
central and spin orbit functions are again taken of Gaussian shape.

The strength and range of the Gaussians are fitted to reproduce the available experimental information for $^7$Li \cite{Tilley2002}.  The $^7$Li spectrum is formed by
negative parity states. Since the $\alpha$ and $t$ both have positive parity (0$^+$ and $\frac{1}{2}^+$), only odd values for the relative orbital angular momentum are 
relevant for the description of the lowest states. In particular $\ell=1$ and $\ell=3$ will be considered. To make sure that positive parity states are pushed away, a repulsive $s$-wave interaction will be considered, whereas for $\ell=2$ the repulsive centrifugal barrier does the job, and the $d$-wave nuclear potential can be assumed equal to zero.

For $\ell=1$, the coupling with the $\frac{1}{2}^+$-triton state gives rise to $\frac{3}{2}^-$ and $\frac{1}{2}^-$ states in $^7$Li. These two states are observed experimentally below the 
$\alpha$-triton threshold with separation energies of 2.467 MeV and 1.989 MeV, respectively \cite{Tilley2002}. For the central and spin-orbit Gaussians we choose a range of 4 fm, and the
two strengths are fitted to reproduce the two separation energies mentioned above.  The choice of this range is somewhat arbitrary, but no relevant changes have been
observed in the final results when modifying this range within reasonable limits. 

For $\ell=3$, the coupling with the triton angular momentum gives rise to a $\frac{5}{2}^-$ and a $\frac{7}{2}^-$ state. Experimentally there is a 
$\frac{7}{2}^-$ resonance with energy 2.185 MeV
(above the $\alpha$-$t$ threshold) with a width of 69 keV. For the $\frac{5}{2}^-$ state, two resonances are observed experimentally. We choose the lowest one, with energy
4.137 MeV and width 0.918 MeV, as the $f_{5/2}$ partner of the $f_{7/2}$ state at 2.185 MeV. For this resonance only the $\alpha$-$t$ decay channel is open, whereas for
the higher $\frac{5}{2}^-$ state the decay through the $^6$Li+$n$ channel is also possible, and the description of this $^7$Li state assuming an $\alpha$-$t$ configuration
is very likely not appropriate. To fit the two resonance energies and the two widths, we take the 
$\ell=3$ potential as a sum of two Gaussians with the two strengths and two ranges as fitting parameters of the four experimental data.

In the lower part of Table~\ref{tab2} we collect all the parameters for the $\alpha$-triton central and spin-orbit Gaussian potentials for the different partial waves. For $\ell=3$ we provide the two
strengths and two ranges corresponding to the two Gaussians whose sum gives the potential.

\subsection{$\alpha$+$d$+$^2n$ configuration}
\label{secad2n}

\begin{table}[t!]
\begin{tabular}{|ccc|c|}
\hline
  & \multicolumn{3}{c|}{$\alpha$-deuteron interaction} \\
\hline
   &
\multicolumn{2}{c|}{$V_c^{(\ell)}$ } &
\multicolumn{1}{c|}{$V_\mathrm{so}^{(\ell)}$ } \\
\hline
$\ell\,\,\,\,$ & $0$ &  $2$   & $2$ \\
\hline
$S\,\,\,\,$  & $-11.16$  & $-40.81$, $-35.68$  & $-13.62$, $17.84$ \\
$b\,\,\,\,$  & 3.5   & 2.5, 2.0    & 2.5, 2.0 \\ \hline
\multicolumn{4}{c}{ } \\ \hline
& \multicolumn{3}{c|}{$\alpha$-dineutron interaction} \\ 
\hline
   &
\multicolumn{2}{c|}{$V_c^{(\ell)}$ } &
\multicolumn{1}{c|}{$V_\mathrm{so}^{(\ell)}$ } \\
\hline
$\ell\,\,\,\,$ & $0$   & $2$   & $2$\\
\hline
$S\,\,\,\, $  & $-20.37$  &  $-63.35$   & --- \\ 
$b\,\,\,\,$  & 1.8    &   2.50 & --- \\
\hline
\end{tabular}
\caption{Strength $S$ and range $b$ of the Gaussian central ($V_c^{(\ell)}$) and
spin-orbit ($V_\mathrm{so}^{(\ell)}$) potentials in Eq. (\ref{pot2bb}) for the $\alpha$-deuteron (upper part) and
$\alpha$-dineutron  (lower part) systems.  For the $\alpha$-deuteron case and  $\ell=2$ the potential is given as the sum of two Gaussians.
The strengths are given in MeV and the ranges in fm.}
\label{tab3}
\end{table}

\paragraph{The $\alpha$-deuteron potential:}
For $s$-waves, the $\alpha$-$d$ interaction is constructed to reproduce the deuteron separation energy in $^{6}$Li, 1.474 MeV \cite{Tilley2002}. This is done with just one 
Gaussian potential with strength $-11.16$ MeV and range 3.5 fm.  Again, as in the $\alpha$-triton potential, a reasonable change of the range in the $s$-wave potential
does not significantly modify the results.

For $d$-waves, we construct a potential of the form given in Eq.(\ref{pot2bb}), where the central and spin-orbit terms are adjusted to reproduce the
experimental energy and width of the 3$^+$, $(E_R,\Gamma_R)$=(0.712,0.024) MeV, and $1^+$, $(E_R,\Gamma_R)$=(4.18,1.5) MeV, resonances in $^6$Li, \cite{Tilley2002}.
This is achieved with a central potential given as sum of two Gaussians with strengths $-40.81$ MeV and $-35.68$ MeV, and ranges 2.5 fm and 2.0 fm, respectively.
For the spin-orbit term, we also have a two-Gaussian potential with strengths   $-13.62$ MeV and $17.84$ MeV, and ranges 2.5 fm and 2.0 fm. This $d$-wave potential 
gives rise to a $2^+$ resonance in $^6$Li with  $(E_R,\Gamma_R)$=(2.71,0.75) MeV, in reasonable agreement with the experimental $2^+$ state
with $(E_R,\Gamma_R)$=(2.84,1.30) MeV reported in Ref.\cite{Tilley2002}.

The negative parity states in $^6$Li are very high in energy. The components with odd relative angular momentum between the $\alpha$ and the deuteron are playing a very 
minor role. In fact, we have checked that the results do not really change when the potential for these components is assumed to be zero, or somewhat attractive (for instance
the same as for $s$-waves when $\ell=1$ and the same as for $d$-waves when $\ell=3$), or even excluded from the calculation.

Obviously, these nuclear potentials must in all cases be completed with the Coulomb repulsion between the $\alpha$-particle and the deuteron.

\paragraph{The $\alpha$-dineutron potential:}
From standard $^6$He ($\alpha$+$n$+$n$) three-body calculations, it is known that the $0^+$
ground state in $^6$He is by far dominated by the component with a relative $s$-wave between the two neutrons and
a relative $s$-wave between the $\alpha$-particle and the dineutron center of mass \cite{Garrido1999b}. This is consistent
with the treatment of the dineutron as a spin zero particle, since a relative $s$-wave between the two neutrons necessarily
implies spin equal to zero due to the antisymmetry requirement.
Therefore, the $\alpha$-$^2n$ $s$-wave  potential is constructed such that the experimental two-neutron separation
energy, $-0.975$ MeV \cite{Brodeur2012}, is reproduced. This is done with Gaussian strength $S=-20.37$~MeV and range $b=1.8$~fm.

The $^6$He nucleus has a $2^+$ resonance located at 0.83 MeV
above the two-neutron threshold with a width of 0.113 MeV
\cite{Tilley2002}.  Three-body calculations of this resonance
\cite{Fedorov2003} show that the component with relative $s$-wave
between the two-neutrons and relative $d$-wave between the
$\alpha$-particle and the dineutron center of mass is dominating.
However, other non-negligible components also contribute, like
relative $d$-wave between the two neutrons and an $s$-wave between the
$\alpha$-particle and the dineutron.  We take then this $2^+$
experimental information to determine the strength and range of the
$d$-wave $\alpha$-dineutron interaction, such that the experimental energy and
width of the $2^+$ resonance in $^6$He is reproduced. This is done
with a Gaussian potential with strength and range $S=-63.35$~MeV and
$b=2.50$~fm, respectively.

As in the $\alpha$-$d$ potential, a $p$-wave interaction between the $\alpha$-particle and the dineutron should in principle be considered. 
However, since we are assuming that the dineutron is a spin-zero  particle, this interaction would be determined by the energy
of the $1^-$ state in $^6$He. Since this state has not been found in $^6$He, or in any case it is very high in energy,
the components with relative $p$-wave between the $\alpha$ and the dineutron can not play  a crucial role.  Again,
the differences in the final results when putting this potential equal to zero, putting it equal to the $s$-wave potential, or
even excluding the $p$-wave components from the calculation, are very limited. The same happens with the $f$-wave potential.

\paragraph{The deuteron-dineutron potential:}
This effective potential is certainly rather uncertain.  As mentioned above, $^4$H has two sets of negative parity resonances, $\{0^-,1^-\}$ and $\{1^-,2^-\}$, that can be nicely
described as a $t$-$n$ system with the neutron in a relative $p_{1/2}$ or $p_{3/2}$ wave. In fact, these resonances are all below the deuteron-neutron
threshold of the triton.  The consequence is that we can not use the experimental information about the $^4$H resonances in order to estimate the interaction between 
the deuteron and the dineutron.

In any case, we certainly have the constraint derived from the fact
that the $d$-$^2n$ system has to be unbound, which of course limits
the attraction provided by the potential.  For this reason we use a
relatively weak Gaussian, where a range of $2.5$~fm and a strength of
about $-8$~MeV is the threshold for binding $d$-$^2n$ in an $s$-wave.
We choose then a little less attractive strength of $-6.5$~MeV for the interaction,
which corresponds to an $s$-wave scattering length of about $a_{d,^2n}
\simeq -7.5$~fm and a virtual state of energy $\hbar^2/(2m a^2_{d,^2n})
\simeq 0.4$~MeV, where $m$ denotes the nucleon mass.

\section{The three-body method}
\label{sec3}

The three-body, $\alpha$+$t$+$n$ and $\alpha$+$d$+$^2n$, wave functions will be computed after solving the Faddeev equations in coordinate space by means of the hyperspherical adiabatic
expansion method. The procedure is described in detail in Ref.~\cite{Nielsen2001}, where the three-body bound states are obtained by looking for the solutions of the Faddeev equations with an exponential asymptotic fall-off of the radial wave functions.

The basics of the method relies on the use of the Jacobi coordinates, $\bm{x}$ and $\bm{y}$, to describe the system, where, except for some mass factors, the $\bm{x}$-vector
connects two of the particles, and $\bm{y}$ connects the third particle and the center of mass of the other two. Obviously, depending on the choice made for the two particles connected
by $\bm{x}$, three different sets of Jacobi coordinates are possible. The total wave function is then written as a sum of three components, each of them depending on
each of the three possible sets of Jacobi coordinates, and each of them being the solution of each of the three Faddeev equations, see Ref.~\cite{Nielsen2001}.

When solving the three-body problem, it is also necessary to specify the components included in the calculation.  We shall characterize them with
the quantum numbers  that are not mixed by the potential in Eq.(\ref{pot2ba}), i.e., $\{\ell_x,j_n,j_x,\ell_y,j_y\}$, where $\ell_x$ and $\ell_y$ are the
relative orbital angular momenta between the two particles connected by $\bm{x}$, and between the third particle and the center of mass of the first two,
respectively.  With this choice $j_x$ is nothing
but the angular momentum of the two-body system formed by the two particles connected by $\bm{x}$. This angular momentum, $j_x$, couples 
to the angular momentum $j_y$, which results from the coupling between $\ell_y$ and the spin of the third particle, to produce the total angular momentum $J$.

In the $\alpha$+$t$+$n$ configuration, when $\bm{x}$ connects the triton and the neutron, as given below
Eq.(\ref{pot2ba}), $j_n$ results from the coupling between $\ell_x$ and the neutron spin, which after coupling to the spin of the triton gives rise to $j_x$. 
For the other two Jacobi sets, due to the zero spin of the $\alpha$-particle, the two-body potential reduces to the form in Eq.(\ref{pot2bb}), and the quantum number
$j_n$ is not necessary (it is always equal to either $\ell_x$ or $j_x$, depending on the coupling choice).  The same happens in the
$\alpha$+$d$+$^2n$ configuration, where at least one of the two particles connected by $\bm{x}$  has zero spin.

It is well known that three-body potentials play an important role in the description of three-body systems. In this work, when needed, we will consider a Gaussian three-body
force,
\begin{equation} 
V_\mathrm{3b}=S_\mathrm{3b}e^{-\rho^2/\rho_0^2},
\label{3bdf}
\end{equation}  
which depends only on the hyperradius $\rho=\sqrt{x^2+y^2}$.  Since
$\rho$ is invariant with respect to choice of Jacobi set, this
three-body potential does not prefer any specific structure. The choice of interaction is only lifting or lowering the energy, while essentially leaving
the three-body structure untouched. The three-body potential also takes care of all those effects that, due to the
intrinsic structure of the clusters, go beyond the three-body level. An example could be the inner structure of the dineutron,
which could be slightly different when two systems are interacting instead of just one. This is rather unpredictable in the present
situation, also because it could depend on the structure following the quantum numbers, which can result in different three-body
potentials for different three-body states.  In all the results shown in the following sections we have taken
$\rho_0=4.5$~fm.  For a larger (smaller) $\rho_0$, the strength would have to be
reduced (increased) to maintain the same effect.  The Faddeev equations are not a
variational method, and the three-body strength can be either
attractive or repulsive, i.e. with both signs of $S_\mathrm{3b}$.

The three-body method was extended to compute resonances by combining
the hyperspherical adiabatic expansion \cite{Nielsen2001} with complex
scaling \cite{Moiseyev1998}.  Resonances appear then as states with
complex energy, whose wave function behaves as a bound state, i.e., the
solutions fall off exponentially at large distances
\cite{Fedorov2003}.  The real part of the energy, $E_R$, gives the
resonance energy, whereas the imaginary part, $E_I$, gives the
resonance width, $\Gamma_R$, by means of the well known relation
$\Gamma_R=-2E_I$.

In this work the radial wave functions will be obtained by imposing a box boundary condition, which automatically discretizes the continuum spectrum. The continuum states are then located along different cuts in the fourth quadrant of the complex energy plane. The cuts are rotated by an angle $2\theta$ (where $\theta$ is the angle chosen for the complex rotation) and they correspond to the different possible continuum structures.
The cut starting at the origin of the energy plane represents genuine three-body continuum states, whereas the cuts starting a the different possible bound or resonant two-body
energies represent structures, where two of the particles populate the corresponding two-body state, bound or resonant, with the third particle in the continuum. If the complex
scaling angle, $\theta$ , is big enough, the three-body resonances can then be identified as the states whose energy and width, which are $\theta$-independent, are located as points
separated from the different cuts in the energy plane \cite{Moiseyev1998}.

\section{$^8$Li as an $\alpha$+$t$+$n$ system}
\label{sec:atn}

\begin{figure}[t!]
\centering
\includegraphics[width=8cm]{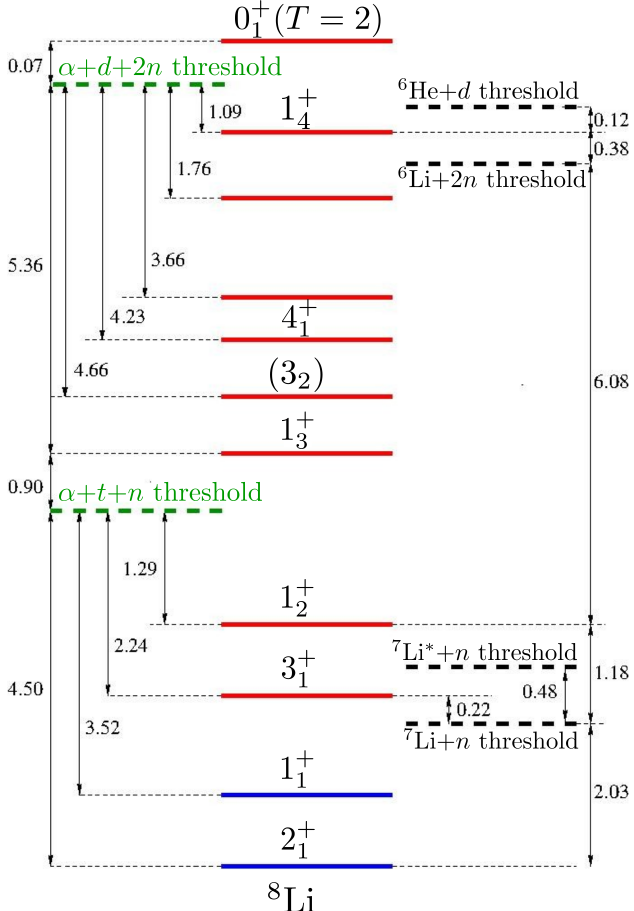}
\caption{Scheme of the experimentally known bound (blue bars) and low-lying resonances (red bars) in $^8$Li. The thick dashed-bars indicate the two-body 
	(black) and three-body (green) thresholds. The energies, in MeV, have been taken from Ref.~\cite{Tilley2004}. To make the figure clean the energy separation between the states in not proportional. }
\label{fig:sch}
\end{figure}

To facilitate the understanding of the following discussions, we show in Fig.~\ref{fig:sch} a scheme of the experimentally known $^8$Li states relevant for this work. The bound 
states are shown by the blue bars, whereas the red bars are $^8$Li resonances. The dashed bars in the right and left parts of the figure indicate the position of the possible two-body
and three-body  thresholds,  respectively. The energies are given in MeV, and they have been taken from Ref.~\cite{Tilley2004}. To make the figure clean the energy separation
between the states is not proportional to the real scale.

Note that all the experimentally known states in the figure correspond
to total isospin equal to 1 \cite{Tilley2004}.  The only exception is
the highest $0^+$ resonance, which is the isobaric analogue of the
ground state of $^8$He with corresponding total isospin of $2$.  As
mentioned in the introduction, such a state can not be described by
any of the two three-body structures considered here, and it is
therefore out of our scope.
It would rather have to be a structure like $\alpha $+$^2n$+$np$, where $np$ is
in an isospin 1 state, the isobaric analogue of $^2n$.

\subsection{Bound states}

\paragraph{Energy and structure of the $J^\pi=2^+,1^+$ states:}
The computed three-body ground state has {\bf $J^\pi=2^+$}. Using the two-body interactions described in Section~\ref{sec:2bda}, we get a 
separation energy of $-4.65$ MeV, which is in fair agreement with the experimental value
of $-4.49920$ MeV given in \cite{Tilley2004}. Therefore, a very small repulsive three-body force is enough
to fit the experimental value (we use the Gaussian three-body force given in Eq.(\ref{3bdf}) with $S_\mathrm{3b}=0.51$ MeV).  When this is done we obtain a root-mean-square 
(rms) radius of 2.53 fm (assuming rms radii for the triton and the $\alpha$-particle equal to 1.7 fm), which is in agreement with the experimental value of $2.51\pm 0.03$ fm
reported in \cite{Tanihata1985}, although somewhat bigger than the more recent experimental result of $2.39\pm 0.05$ fm reported in \cite{Fan2014}.

\begin{table}[t!]
\begin{center}
  \begin{tabular}{|ccccc|ccccc|ccccc|}
\hline    
\multicolumn{15}{|c|}{2$^+_1$ at $-4.50$ MeV (ground state), $S_\mathrm{3b}=0.51$ MeV}  \\ \hline
\multicolumn{5}{|c|}{$^4$H($t,n$)+$\alpha$}    &  \multicolumn{5}{c|}{$^5$He($\alpha,n$)+$t$}  &  \multicolumn{5}{c|}{$^7$Li($\alpha,t$)+$n$}   \\
\hline    
  $\ell_x$   &  $j_x$  &  $\ell_y$ & $j_y$ &  weight  &  $\ell_x$  &  $j_x$ &  $\ell_y$ & $j_y$ &  weight    &  $\ell_x$  &  $j_x$ &  $\ell_y$ & $j_y$ &  weight  \\ \hline
     1             &   1    &      1          &      1    &    19.8   & 1  & $\frac{1}{2}$  &  1 & $\frac{3}{2}$  & 10.0   &  1  &  $\frac{1}{2}$    & 1   &  $\frac{3}{2}$  &  17.9 \\
     1            &    2    &      1          &      1    &    74.1   &  1  & $\frac{3}{2}$     & 1   & $\frac{1}{2}$  & 17.5 &  1  &  $\frac{3}{2}$  &  1  &  $\frac{1}{2}$  &  9.7\\
     2             &   3    &      2          &      2    &    2.0     &   1  &  $\frac{3}{2}$    & 1  &  $\frac{3}{2}$ &  70.6 &  1   & $\frac{3}{2}$     & 1   &   $\frac{3}{2}$ & 70.4 \\  \hline
\multicolumn{15}{c}{ } \\
 \hline
\multicolumn{15}{|c|}{1$^+_1$ at $-3.52$ MeV (0.98 MeV exc. energy), $S_\mathrm{3b}=-0.38$ MeV}  \\ \hline
\multicolumn{5}{|c|}{$^4$H($t,n$)+$\alpha$}    &  \multicolumn{5}{|c|}{$^5$He($\alpha,n$)+$t$}  &  \multicolumn{5}{|c|}{$^7$Li($\alpha,t$)+$n$}   \\
\hline
  $\ell_x$   &  $j_x$  &  $\ell_y$ & $j_y$ &  weight  &  $\ell_x$  &  $j_x$ &  $\ell_y$ & $j_y$ &  weight    &  $\ell_x$  &  $j_x$  &  $\ell_y$ & $j_y$ &  weight  \\ \hline
     1             &   0    &      1          &      1    &    3.7   & 1  & $\frac{1}{2}$  &  1 & $\frac{1}{2}$  & 9.5   &  1  &  $\frac{1}{2}$    & 1   &  $\frac{1}{2}$  &  8.8 \\
     1             &   1    &      1          &      1    &    82.5     &  1  & $\frac{3}{2}$     & 1   & $\frac{1}{2}$  & 37.7 &  1  &  $\frac{1}{2}$  &  1  &  $\frac{3}{2}$  &  37.4\\
     1             &   2    &      1          &      1    &    7.4      &   1  &  $\frac{3}{2}$    & 1  &  $\frac{3}{2}$ &  50.8 &  1   & $\frac{3}{2}$     & 1   &   $\frac{3}{2}$ & 50.7 \\
     2             &   2    &      2          &      2    &    2.4     &      &   &    &   &  &  2  & $\frac{5}{2}$   &  2  &  $\frac{5}{2}$  &  1.1  \\     \hline
\end{tabular}
\end{center}
\caption{For each of the  three Jacobi sets, components contributing with more than 1\% to the $2^+_1$ (upper part) and $1^+_1$ (lower part) bound state wave 
functions in $^8$Li. The first line in each of the tables gives the computed separation energy referred to the three-body threshold, the corresponding excitation
energy, and the strength, $S_\mathrm{3b}$, used in three-body potential in Eq.(\ref{3bdf}).}
\label{tab2plus}
\end{table}

The calculation has been performed including up to six adiabatic
channels, although the lowest alone provides more than 99\% of the
norm. All the partial wave components with $\ell_x,\ell_y\leq 3$ in
the three possible sets of Jacobi coordinates are included.

In the upper part of Table~\ref{tab2plus} we specify, for each Jacobi
set, those components that contribute with more than 1\% to the total
norm of the bound state.  The $j_x$ quantum number is the angular
momentum of the two-body system (bound or in the continuum) formed by
the two particles connected by the $\bm{x}$ Jacobi coordinate, i.e,
$^4$H, $^5$He, or $^7$Li, which are shown in the left, central, and
right parts of the table, respectively.  We shall refer to them as 
first, second, and third Jacobi sets, respectively.

We see that, on each of the
three Jacobi sets, the computed 2$^+$ ground state in $^8$Li is
dominated by $^4$H in the $2^-$ state plus $\alpha$ ($\sim 74$\%),
$^5$He in the $3/2^-$ state plus triton ($\sim 88$\% in total), and
$^7$Li in the $3/2^-$ plus neutron ($\sim 80$\% in total),
respectively. In other words, the three-body ground state is dominated
by structures with the two-body system with spin and parity
corresponding to its ground state plus the third particle in a
$p$-wave.  These structures describe precisely the same three-body
system in the different coordinate systems, and are therefore only
different ways of visualizing an identical quantum state.

The bound excited {\bf $1^+$} state is found numerically with an energy of  $3.43$ MeV below the $\alpha$+$t$+$n$ threshold,
which is very close to the experimental value of $-3.5184$ MeV \cite{Tilley2004}, see Fig.~\ref{fig:sch}. Therefore, also in this
case, a small three body force is sufficient to fit the experiment, i.e. $S_\mathrm{3b}=-0.38$ MeV in Eq.(\ref{3bdf}). The computed rms radius is 2.76 fm. 
In this case the first adiabatic channel gives 90\% of the norm, 
and the three lowest channels are needed to approach 99\% of the norm. The calculation has again been made 
including six adiabatic terms and all the components with $\ell_x,\ell_y \leq  3$.

In the lower part of Table~\ref{tab2plus} we also give the weights of
the components contributing with more than 1\% to the norm of the
$1^+$ wave function.  We can see that in this case the components with
$^4$H and $^7$Li in an excited state contribute substantially.  In the
first Jacobi set (left part), about 83\% of the wave function is $^4$H
in a $1^-$ state, and in the third Jacobi set (right part), about 46\%
of the wave function is $^7$Li in a $1/2^-$ state.  The second Jacobi set
(middle part) show that 20\% is moved from the triton in a $3/2^-$ to a
$1/2^-$ state, while leaving $^5$He in the $3/2^-$ state. Again the
third particle is in all cases dominantly in a $p$-wave.  All these
descriptions are still of the same $1^+$ state.

\begin{table}[t!]
  \begin{tabular}{|c|cc|cc|cc|}
 \hline   
 & \multicolumn{2}{c|}{$^4$H($t,n$)+$\alpha$    }    &  \multicolumn{2}{c|}{$^5$He($\alpha,n$)+$t$ }  & \multicolumn{2}{c|}{$^7$Li($\alpha,t$)+$n$ } \\ \hline  $J^\pi$ &
$\langle r^2_{tn} \rangle^{\frac{1}{2}}$  &  $\langle r^2_{\alpha ^4\mathrm{H}} \rangle^{\frac{1}{2}}$  & $\langle r^2_{\alpha n} \rangle^{\frac{1}{2}}$  & $\langle r^2_{t ^5\mathrm{He}} \rangle^{\frac{1}{2}}$  & $\langle r^2_{\alpha t} \rangle^{\frac{1}{2}}$  & $\langle r^2_{n ^7\mathrm{Li}} \rangle^{\frac{1}{2}}$   \\ \hline
$2^+_1$ & 4.47 &  2.85  &  4.18    & 3.04    &  3.16  &  4.03   \\
$1^+_1$ &   5.16 & 3.19 &  4.80  &  3.43  &  3.54  &  4.63 \\ \hline
\end{tabular}
  \caption{Root-mean-square radii, $\langle
    r_x^2\rangle^{\frac{1}{2}}$ and $\langle
    r_y^2\rangle^{\frac{1}{2}}$, for the bound $2^+_1$ and $1^+_1$
    states in the three Jacobi sets.  These radii are given in fm and
    defined as the physical lengths corresponding to the $x$ and $y$
    coordinates where the the mass factors are included. }
\label{tabr2plus}
\end{table}

\paragraph{Spatial distributions of the $J^\pi=2^+,1^+$ states:}
The geometry of the system can be seen in Table~\ref{tabr2plus}, where we give the $r_x$ and $r_y$ rms radii in all the three Jacobi sets for the $2^+$ and $1^+$ bound states. From the radii we can infer that the geometry of both states resembles an isosceles triangle with the short side between the alpha and the triton, separated by about 3.2 fm and 3.6 fm
in the ground and excited states, respectively. The neutron, located on top of the isosceles triangle, is about 4.3 fm and 5.0 fm away from the other two particles 
for each of the two states.

\begin{figure}[t!]
\includegraphics[width=\columnwidth]{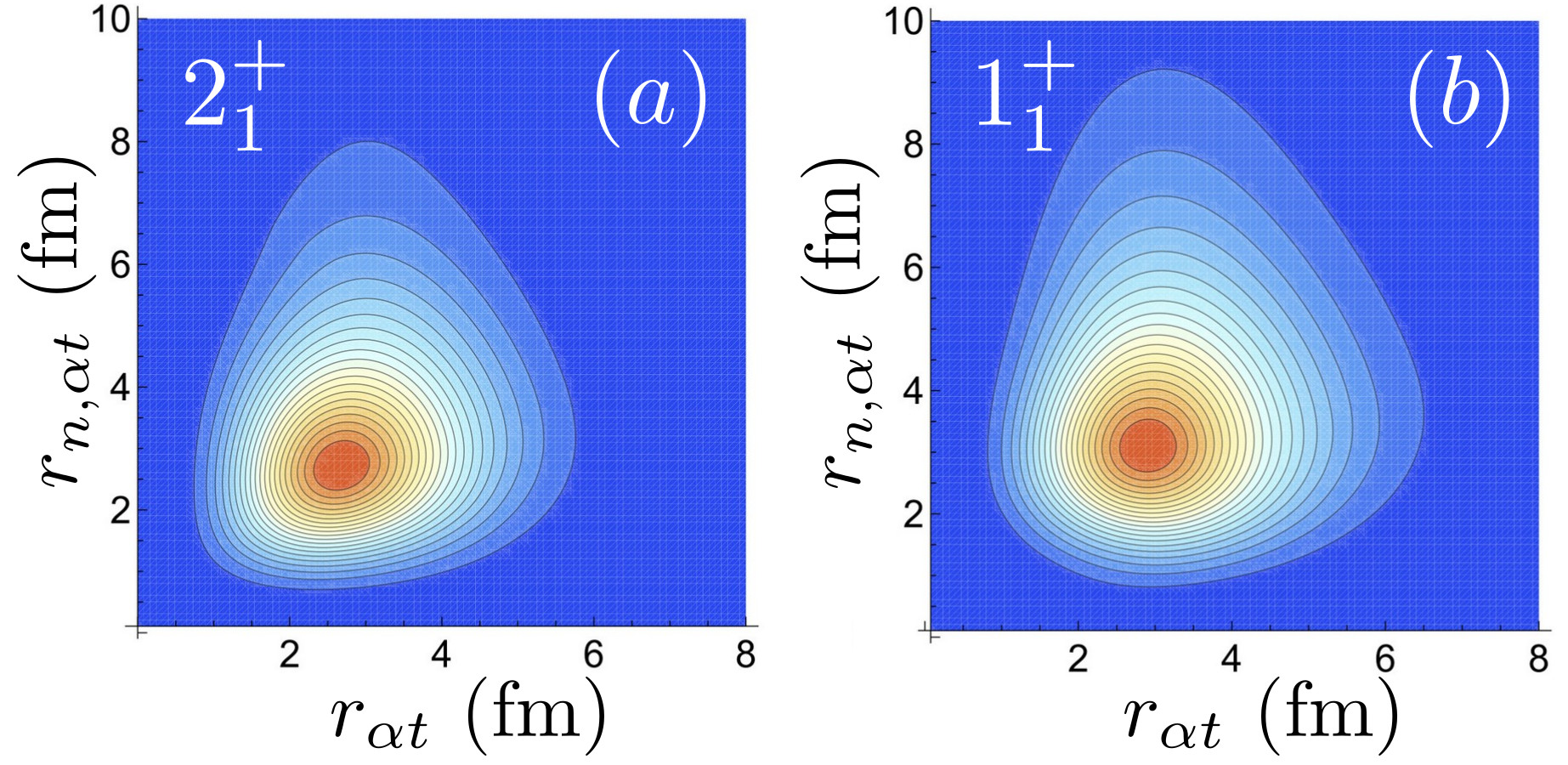}
\caption{Angle integrated three-body wave functions, Eq.(\ref{intwf}), for the 2$^+_1$  (a) and $1^+_1$ (b) bound states in $^8$Li in
the Jacobi set where $\bm{x}$ connects the $\alpha$ and the triton.}
\label{fig1}
\end{figure}

In Fig.~\ref{fig1} we plot
\begin{equation}
F(r_{\alpha t},r_ {n,\alpha t´})=\int d\Omega_{\alpha t} d\Omega_{n,\alpha t} r_{\alpha t}^2 r_{n,\alpha t}^2 |\Psi(\bm{r}_{\alpha t},\bm{r}_{n,\alpha t})|^2,
\label{intwf}
\end{equation}
which visualizes the probability functions for both the $2^+$
(Fig.~\ref{fig1}a) and the 1$^+$ (Fig.~\ref{fig1}b) states.  The
distributions are more compact in the $r_{\alpha t}$, while the
neutron is extending further away from the $\alpha$-triton center of
mass.  Nevertheless, both states have pretty much the same geometry,
but are spatially more extended in the $1^+$ case, as expected due to the
smaller separation energy.

With these $1^+$ and $2^+$ wave functions, we obtain a
$B(E2,2^+_1\rightarrow 1^+_1)$ transition strength that ranges from 0.46
$e^2$fm$^4$ to 0.61 $e^2$fm$^4$ for an effective neutron charge
varying from zero to $0.3e$, which was the value used in
\cite{Romero2008a} to reproduce the experimental value of
$B(E1, 0^+_1\rightarrow 1^-)$ in $^{12}$Be.  The value obtained here
for the $B(E2,2^+_1\rightarrow 1^+_1)$ strength is not far from the 
0.83 $e^2$fm$^4$ obtained in a Quantum Montecarlo calculation
\cite{Pastore2013}, and even from the ab initio estimate, from 1 
to 2 $e^2$fm$^4$, given in \cite{Caprio2022}.  However, these calculated values are very far
from the anomalously large experimental strength of $55\pm 15$
$e^2$fm$^4$ given in \cite{Brown1991}, although reduced to
$19^{+7}_{-6}$ $e^2$fm$^4$ in a recent measurement
\cite{Henderson2024}.

Furthermore, our computed static quadrupole moment for the
$2^+$ state ranges from 2.0 $e$fm$^2$ to 2.7 $e$fm$^2$, which compares
reasonably well with the values ranging from 3.0 $e$fm$^2$ to 3.3
$e$fm$^2$ given in \cite{Pastore2013}, and as well with the
experimental value of $3.14\pm 0.02$ $e$fm$^2$ given in
\cite{Stone2016}.  For the excited $1^+_1$ state we obtain a quadrupole
moment that ranges from 1.6 $e$fm$^2$ to 2.2 $e$fm$^2$, again
depending on the effective charge used for the neutron.

The difference between theoretical and experimental values of
$B(E2,2^+_1\rightarrow 1^+_1)$ is large, even for the large experimental
uncertainty.  The measured value exceeds transition strengths for a
well deformed rotational nucleus. The obvious problem of a substantial
magnetic dipole contribution seems to be excluded in
\cite{Henderson2024}.

\subsection{Resonances}

Energies above the $\alpha$+$t$+$n$ threshold must necessarily be resonances. In general,
they are more difficult to calculate than bound states, and, as mentioned, in this work we are 
doing it by implementing the complex scaling method \cite{Moiseyev1998} into the hyperspherical
adiabatic expansion \cite{Nielsen2001}.  The expected, and mostly also experimentally found, states
with different angular momentum and parity quantum numbers are
investigated separately in the following subsections.

\subsubsection{The 1$^+$ states.} 
Experimentally, together with the bound $1^+_1$ state described in the previous section, $^8$Li is known to have three $1^+$ resonances.
The lowest one, see Fig.~\ref{fig:sch}, with 3.21 MeV excitation energy,  is 1.29 MeV below the $\alpha$+$t$+$n$ threshold, but 1.18 MeV above the $^7$Li-$n$ threshold. Therefore, this state is 
very likely, to a large extent, a two-body, $^7$Li+$n$, resonance. The width of the resonance is not known very precisely, but as given in Ref.~\cite{Tilley2004}, it is estimated to be around 1 MeV.

\begin{figure}
\includegraphics[width=7.5cm]{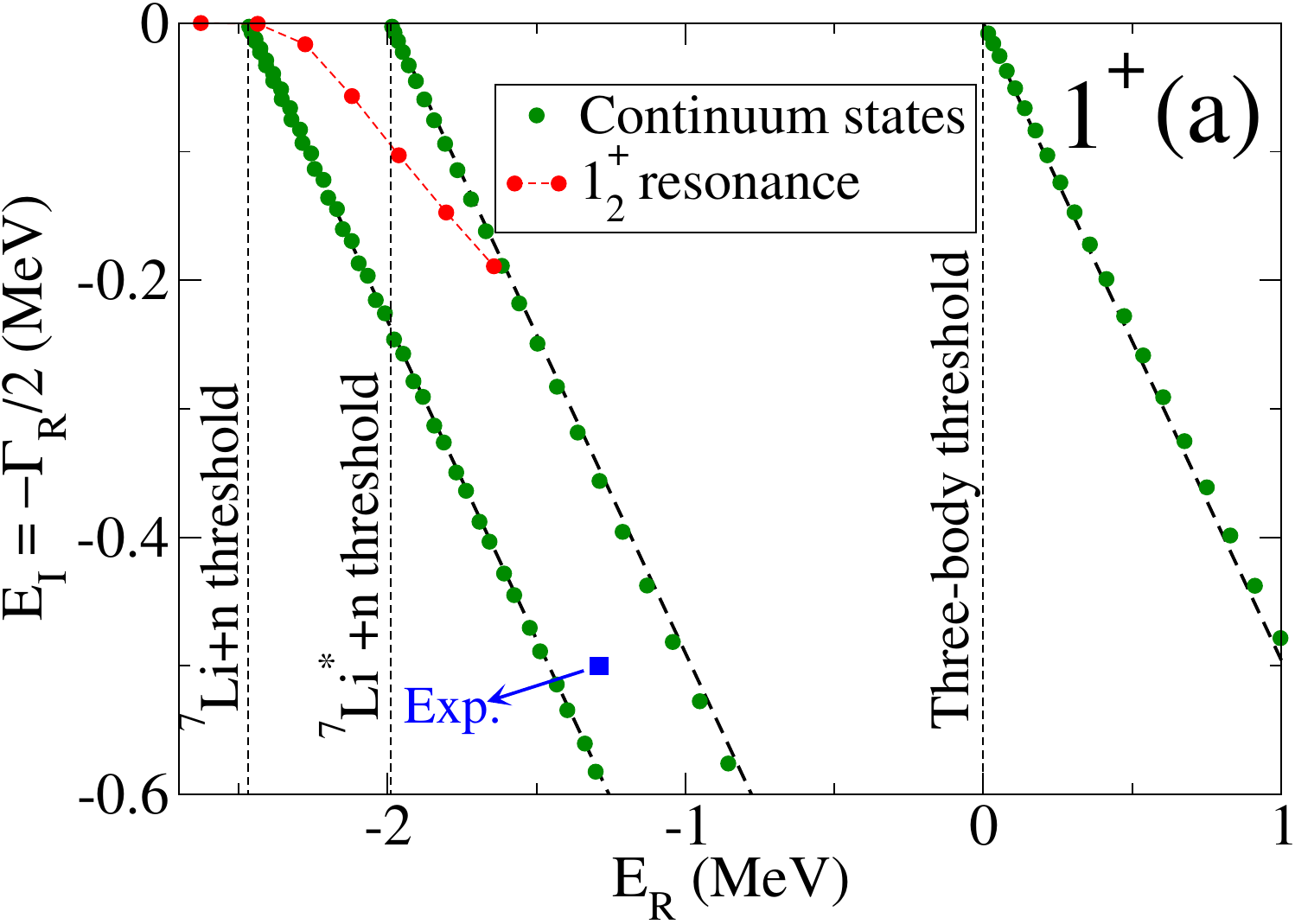}
\includegraphics[width=7.5cm]{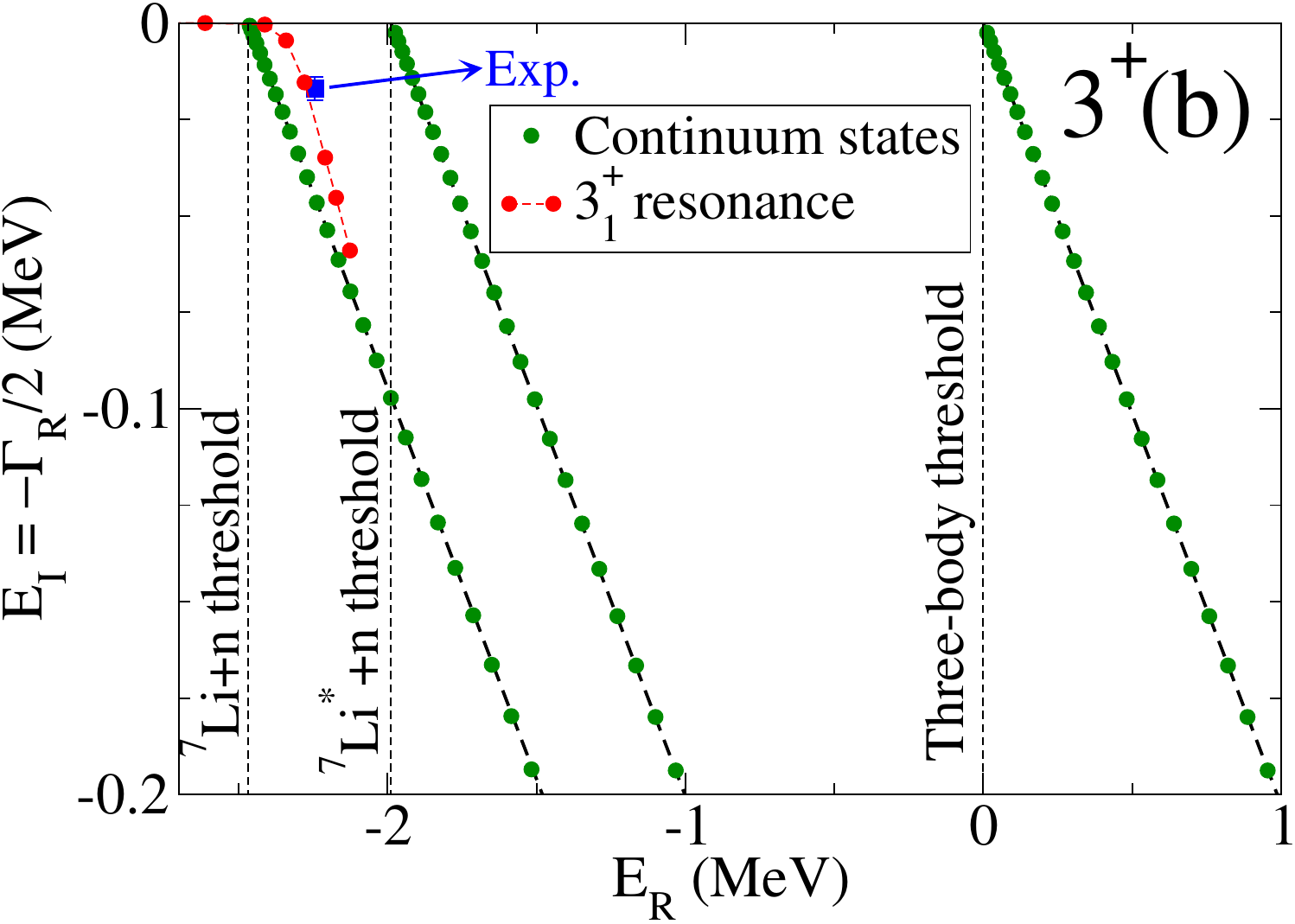}
\includegraphics[width=7.5cm]{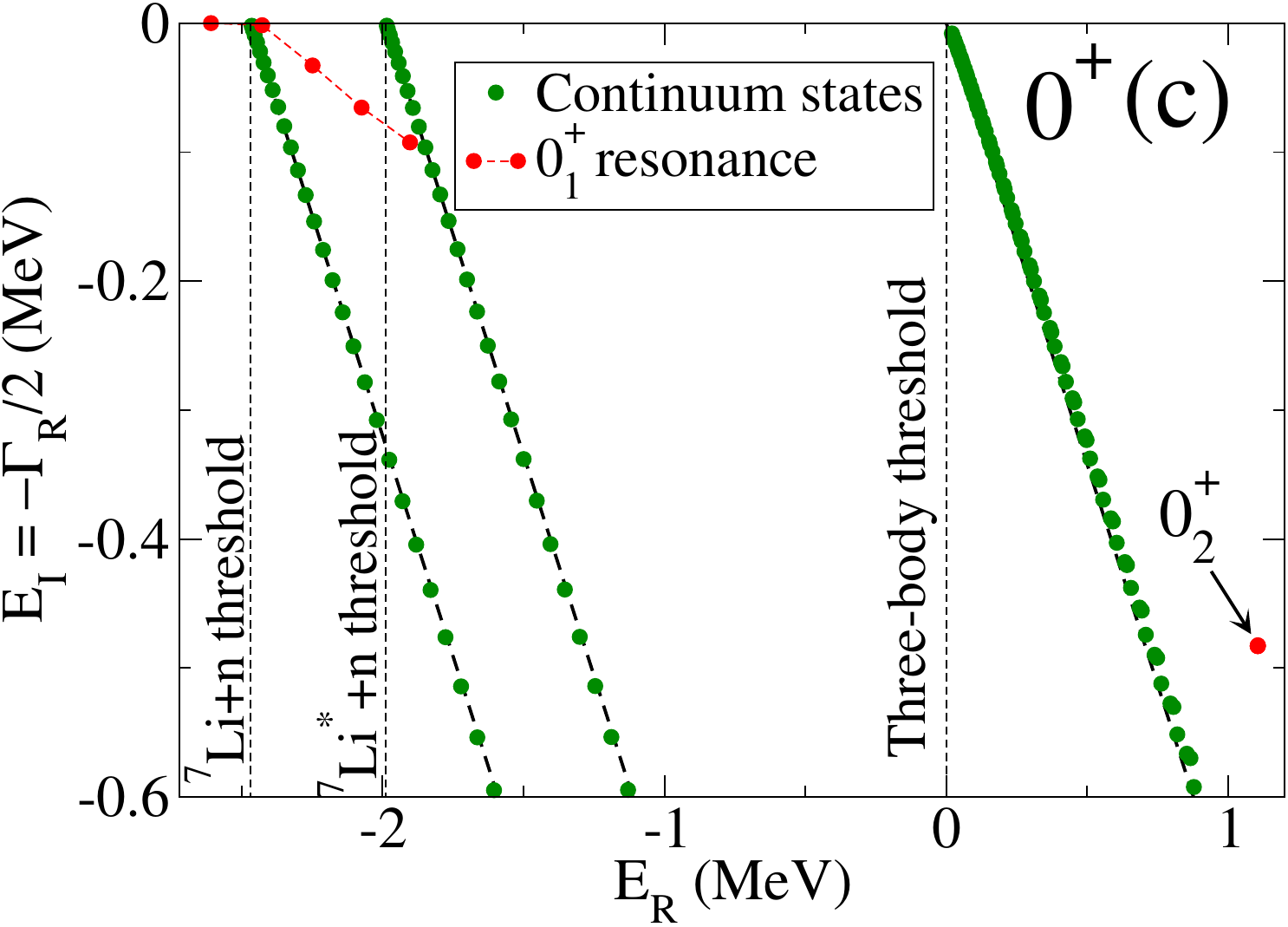}
\caption{Discrete energy spectrum for the 1$^+$ , (a), 3$^+$, (b), and 0$^+$, (c), states obtained with a complex rotation angle $\theta=0.25$ rads,  $\theta=0.10$ rads,
and $\theta=0.30$ rads, respectively. The red dots show the position of the lowest $1^+$, 3$^+$, and $0^+$ resonances when the strength of the three-body force varies from $S_\mathrm{3b}=0$ to   $S_\mathrm{3b}=6$ MeV in (a), from $S_\mathrm{3b}=-12.5$ MeV to   $S_\mathrm{3b}=-10.8$ MeV in (b), and from 
$S_\mathrm{3b}=0$ to   $S_\mathrm{3b}=4$ MeV in (c). When available, the experimental value  \cite{Tilley2004} is indicated by the blue square.}
\label{fig:res1}
\end{figure}

\begin{table}
  \begin{tabular}{|c|ccc|}
     \hline
   & $1^+_2$ &  \multicolumn{2}{c|}{ $^7$Li($\alpha,t$)+$n$ }    \\  \hline
  $S_\mathrm{3b}$ &  $(E_R,\Gamma_R)$   & $\frac{3}{2}^-+n$  & $\frac{1}{2}^-+n$ \\  \hline
  1.0  &   $(-2.44,0.01)$  & 66.5&  33.5\\
  2.0  &   $(-2.28,0.03)$  & 65.8&  34.2  \\
  3.0  & $(-2.12,0.11)$ & 62.1 & 37.9 \\
  4.0 &  $(-1.96,0.21)$ & 58.7& 41.3 \\ 
  5.0  &  $(-1.89,0.29)$& 56.3 & 43.7 \\
  6.0  &  $(-1.64,0.38)$ & 51.8 & 48.2  \\
	Exp. \cite{Tilley2004}&  $(-1.29,\sim 1.0)$ &  &\\ \hline
\multicolumn{4}{c}{ } \\  \hline
   &  $3^+_1$ &  \multicolumn{2}{c|}{ $^7$Li($\alpha,t$)+$n$ }    \\  \hline
  $S_\mathrm{3b}$ &  $(E_R,\Gamma_R)$  & $\frac{3}{2}^-+n$  & $\frac{7}{2}^-+n$ \\  \hline
 $-12.5$ &  $(-2.61,0.00)$ &   59.3   &  34.3   \\
 $-12.0$ &  $(-2.41,0.0)$   & 68.1   &  25.6 \\
 $-11.5$ &  $(-2.28,0.03)$  &   71.2 &    23.4   \\
$-11.0$ &  $(-2.17,0.09)$  &   77.9 &     17.9  \\
Exp.  \cite{Tilley2004}&  $(-2.24, 0.03)$ & &\\ \hline
\multicolumn{4}{c}{ } \\  \hline
   & $0^+_1$  & \multicolumn{2}{c|}{ $^7$Li($\alpha,t$)+$n$ }   \\  \hline
 $S_\mathrm{3b}$ &  $(E_R,\Gamma_R)$   & $\frac{3}{2}^-+n$  & $\frac{1}{2}^-+n$ \\  \hline
  1.0  &   $(-2.43,0.00)$  & 56.9&  42.2  \\
  2.0  & $(-2.25,0.07)$ & 54.0 & 45.0 \\
  3.0 & $(-2.07,0.13)$ & 50.2 & 48.8  \\
  4.0 &  $(-1.90,0.19)$ & 46.7& 52.3 \\  \hline
\end{tabular}
\caption{For the $1^+_2$ (upper part), $3^+_1$  (middle part), and $0^+_1$ (lower part) states,  and for different values of the strength ($S_\mathrm{3b}$, in MeV) of the three-body potential, the second column gives the energies, $E_R$, and widths, $\Gamma_R$ (in MeV and relative to the three-body $\alpha$+$t$+$n$ threshold).
The last two columns give the contribution (in \%) to the wave function of the components corresponding, respectively, to the $\alpha$-$t$
system in the $3/2^-$ state and in the $1/2^-$ state (for the $1^+_2$ and $0^+_1$ cases, or in the $3/2^-$ state and in the $7/2^-$ state, for the $3^+_1$ resonance. When available, the  experimental energy and width are also given.}
\label{tabres1}
\end{table}

Although not mentioned when discussing the bound $1^+_1$ state, the
calculation also provided a second bound 1$_2^+$ state, which, without
three-body force, is located at about 2.6 MeV below the three-body
threshold, that is, about only 100 keV below the two-body, $^7$Li+$n$,
threshold.  As a consequence, a small repulsive three-body force is
sufficient to push this state into the resonance sector. This is seen
in Fig.~\ref{fig:res1}a, where we show the discretized complex rotated spectrum
(with $\theta=0.25$ rads) of the $1^+$ states for different values of
the strength of a Gaussian three-body force. In the figure the red
dots show how the $1^+_2$ state moves when the strength of the
three-body potential, Eq.(\ref{3bdf}), changes from $S_\mathrm{3b}=0$
to $S_\mathrm{3b}=6$ MeV. Specific values of the resonance energy,
$E_R$, and width, $\Gamma_R$, for different values of $S_\mathrm{3b}$
are given in the upper part of the second column in Table~\ref{tabres1}.

As the three-body repulsion increases the resonance moves into the fourth quadrant of the energy plane towards higher energies and widths. 
For $S_\mathrm{3b}\approx 4$ MeV the resonance moves above the $^7$Li$^*$+$n$ threshold. In principle, using a sufficiently 
large three-body repulsion, it should be possible to reach the experimental value, which is indicated in the figure by the blue square. However,
the resonance seems to approach the continuum cut associated to $^7$Li in the  bound $\frac{1}{2}^-$ excited state plus a neutron. In fact, when the three-body repulsion
is still increased, the $1^+_2$ resonance disappears in this continuum.  Therefore, in our approach the resonance is at least about 0.35 MeV lower than the experimental value.

\begin{table}[t!]
\begin{center}
\begin{tabular}{|ccccc|ccccc|ccccc|}
 \hline
  \multicolumn{15}{|c|}{1$^+_2$ at $-1.96$ MeV (2.54 MeV exc. energy), $S_\mathrm{3b}=4.0$ MeV}  \\\hline
\multicolumn{5}{|c|}{$^4$H($t,n$)+$\alpha$}    &  \multicolumn{5}{c|}{$^5$He($\alpha,n$)+$t$  }  &  \multicolumn{5}{c|}{ $^7$Li($\alpha,t$)+$n$ }   \\  \hline
  $\ell_x$   &  $j_x$  &  $\ell_y$ & $j_y$ &  weight  &  $\ell_x$  &  $j_x$ &  $\ell_y$ & $j_y$ &  weight    &  $\ell_x$  &  $j_x$ &  $\ell_y$ & $j_y$ &  weight  \\ \hline
     1             &   0    &      1          &      1    &    28.4   &  1  & $\frac{1}{2}$  &  1 & $\frac{1}{2}$  & 1.5   &  1  &   $\frac{1}{2}$   & 1   &  $\frac{1}{2}$  &  41.3 \\
     1             &   1    &      1          &      1    &    34.6   &  1 &  $\frac{1}{2}$ &   1 & $\frac{3}{2}$  & 47.3   &  1  & $\frac{3}{2}$  &   1    &  $\frac{1}{2}$ &  40.7          \\
     1             &   2    &      1          &      1    &    35.4   &  1 &  $\frac{3}{2}$ &   1 & $\frac{1}{2}$  &  39.0  &  1   &   $\frac{3}{2}$&   1   &  $\frac{3}{2}$  &  18.0         \\
     2             &   1    &      2          &      2    &     1.4    &   1&   $\frac{3}{2}$  & 1 &  $\frac{3}{2}$ &  12.0  &      &         &       &      &            \\
   \hline
\end{tabular}
\end{center}
\caption{The same as Table~\ref{tab2plus} for the $1^+_2$ resonance 1.96 MeV below the $\alpha$+$t$+$n$ threshold.}
\label{tab1plusa}
\end{table}

As an illustration of the detailed partial wave contributions, we show in Table~\ref{tab1plusa} the same component contributions as
in Table~\ref{tab2plus}, but for the resonance in Table~\ref{tabres1}
with energy 1.96~MeV below the $\alpha$+$t$+$n$ threshold.  In the
first Jacobi set (left), where $\bm{x}$ connects the triton and the
neutron, the contribution to the norm of the resonance wave function
is distributed among three possible components. Each combination
describing different $^4$H states, but all with the $\alpha$-particle
in a $p$-state.  In the other two Jacobi sets, the wave functions are
almost completely given by $^5$He in the $\frac{1}{2}^-$ and $\frac{3}{2}^-$ states
(second Jacobi set), or $^7$Li in the $\frac{1}{2}^-$ or $\frac{3}{2}^-$ states (third
Jacobi set), and the third particle in the continuum.

It is interesting to note that, when modifying the strength of the
three-body force, $S_\mathrm{3b}$, the weight of different components
in the third Jacobi set changes significantly.  This is shown in the
last two columns in the upper part of Table~\ref{tabres1}, where we
see that when increasing the three-body repulsion, the weight of the
$\frac{3}{2}^-$ components decreases, whereas that of the $\frac{1}{2}^-$ components
increases.  This is consistent with the progressive approach of the
$1_2^+$ resonance to the energy cut of the $^7$Li$^*$+$n$ continuum
states seen in Fig.~\ref{fig:res1}a.

\begin{figure}[t!]
\centering
\includegraphics[width=\columnwidth]{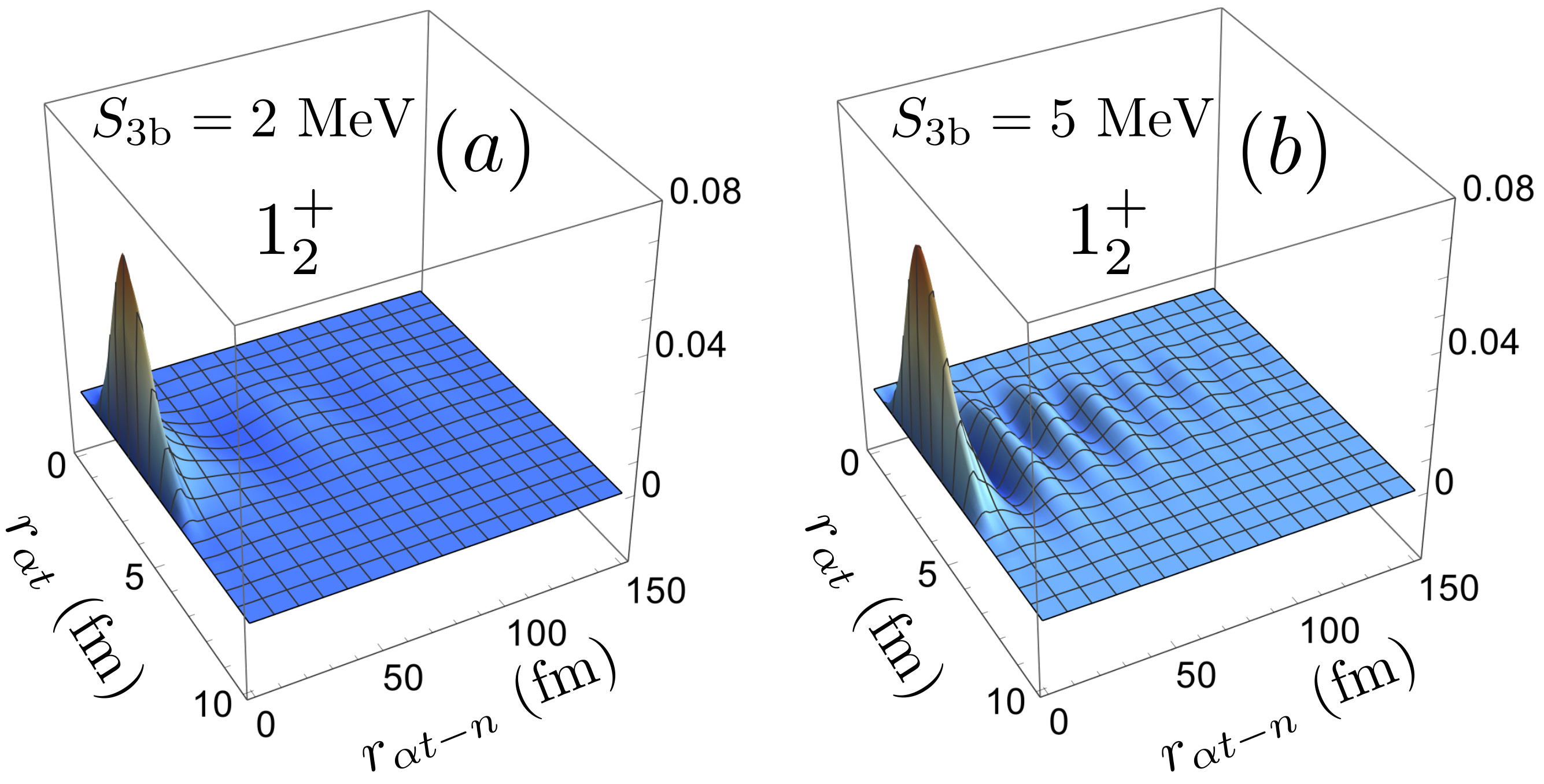}
\caption{For the 1$^+_2$ resonance, square of the complex rotated wave function (integrated over the angles, see Eq.(\ref{wf2})) as a function of the $\alpha$-$t$ distance, $r_{\alpha t}$, and  the distance between the neutron and the $\alpha$-$t$ center of mass, $r_{\alpha t - n}$. The results with two different values of the
the tree-body strength,  $S_\mathrm{3b}$=2 MeV $(a)$ and $S_\mathrm{3b}$=5 MeV $(b)$, are shown.}
\label{fig:3dwf1}
\end{figure}

The approach to the threshold is also reflected in the fact that, when
increasing the three-body repulsion, the $\alpha$-$t$ distance is
rather stable, whereas the neutron progressively moves further apart
from the $\alpha$-$t$ center of mass.  This is shown in
Fig.~\ref{fig:3dwf1}, where we plot the real part of the probability function
\begin{equation}
F(r_{\alpha t}, r_{\alpha t-n})=\int r_{\alpha t}^2 r_{\alpha t-n}^2 \Psi(\bm{r}_{\alpha t},\bm{r}_{\alpha t-n})^2 d\Omega_{\alpha t} d\Omega_{\alpha t-n}   ,           
\label{wf2}
\end{equation}
where $\Psi$ is the complex rotated three-body resonance wave
function, and $\bm{r}_{\alpha t}$ and $\bm{r}_{\alpha t-n}$ are the
relative position vectors between the $\alpha$-particle and the
triton, and between the center of mass of the $\alpha$-$t$ system and
the neutron.  The integration is performed over the angles (collected
in $\Omega_{\alpha t}$ and $\Omega_{\alpha t-n}$) that give the
directions of $\bm{r}_{\alpha t}$ and $\bm{r}_{\alpha t-n}$,
respectively. Note that, contrary to Eq.~(\ref{intwf}), in Eq.~(\ref{wf2})
the wave function is just squared, since this is how wave functions are 
normalized after complex scaling \cite{Moiseyev1998}. The integral
of $F(r_{\alpha t}, r_{\alpha t-n})$ over $r_{\alpha t}$ and 
$r_{\alpha t-n}$ is real and equal to 1.

The figure shows the results for two different values of the
three-body strength, $S_\mathrm{3b}=2$ MeV, Fig.~\ref{fig:3dwf1}a, and
$S_\mathrm{3b}=5$ MeV, Fig.~\ref{fig:3dwf1}b.  As we can see, the
wave functions are pretty much confined in the $r_{\alpha t}$
coordinate. In fact, the expectation value $\langle r_{\alpha t}^2
\rangle^{1/2}$ is rather stable, and takes a value of about $3.7 +i 0.1$ fm.
However, in the $r_{\alpha t-n}$ coordinate, the increasing
$S_\mathrm{3b}$ introduces more and more oscillations in the
long-distance tail.  The wave function becomes more and more
delocalized with oscillations resembling a continuum state, i.e.
corresponding to a large separation and a weak binding between neutron
and the $\alpha$-$t$ center of mass.

For $S_\mathrm{3b}=2$ MeV we get $\langle r_{\alpha t-n}^2
\rangle^{1/2}=5.7 + i 5.1$ fm.  For larger values of $S_\mathrm{3b}$
the calculation of $\langle r_{\alpha t-n} ^2\rangle^{1/2}$ becomes
quickly very delicate, since it requires an accurate calculation of
the wave function at larger and larger distances, which soon makes the
required basis too large for a calculation in a reasonable time.  Note
that, due to the complex scaling transformation, the root-mean-square
distances are actually complex numbers, in such a way that, similarly
to the case of the complex energy, the imaginary part of the
observable can be interpreted as the uncertainty in the observable
measurement (see Ref.~\cite{Moiseyev1998}).

The approach to the continuum of the resonance in Fig.~\ref{fig:res1}a
must be because the necessary confining ``barrier'' in the continuum
is not present in this three-body structure.  The $p$-wave centrifugal
barrier is too small at these energies, and the resonance behavior
with the remarkably large width of about $1$~MeV can not be supported.
Instead, when forced to move up in energy, the excited $^7$Li-state
plus a continuum neutron is approached.

As shown in Fig.~\ref{fig:sch}, two additional $1^+$ resonances have
been found experimentally at 0.90 MeV and 5.17 MeV above the
three-body threshold, respectively.  The width of these resonances are
of about 0.65 MeV and 1.0 MeV. They are therefore  resonances whose argument
is smaller than the complex rotation angle, $\theta=0.25$ rads,
used in the calculation.  However, although the complex rotation
angle used is in principle large enough, these resonances have not
been found numerically.  This could be, because either the resonances
are out of the $\alpha$+$t$+$n$ structure, or the model produces too
wide resonances, such that a larger complex scaling angle is needed to
capture them.

Therefore within the $\alpha$+$t$+$n$ configuration, our calculation
only obtains the lowest $1^+_2$ resonance, and we  find this
structure located at least 0.35 MeV  below the experimental
value (although above the  $^7$Li$^*$-neutron threshold, with
$^7$Li in the excited 1/2$^-$ state).

\subsubsection{The 3$^+$ states.}

The only $3^+$ resonance identified experimentally in $^8$Li is $2.244\pm 0.003$ MeV below the $\alpha$+$t$+$n$ threshold (0.223 MeV above the $^7$Li-$n$ threshold). This is a pretty narrow
resonance, with a width of only $33\pm 6$ keV \cite{Tilley2004}.

In order to get a resonance in this energy region, an attractive three-body force has to be included in the calculations. This is shown in Fig.~\ref{fig:res1}b, where we show
the 3$^+$ states obtained after a calculation using the complex scaling angle $\theta=0.10$ rads. As in Fig.~\ref{fig:res1}a, the red circles show the position of the 
$3^+_1$ resonance for different strength values, $S_\mathrm{3b}$, of the Gaussian three-body force. To be precise, in the figure $S_\mathrm{3b}$ ranges from 
$-12.5$ MeV (which produces a bound state) to $-10.8$ MeV. As we can see, for $S_\mathrm{3b} \approx -11.5$ MeV the experimental value (indicated by the blue square
in the figure) is  well reproduced.  The energy and width, $(E_R,\Gamma_R)$, of the $3^+_1$ state for specific values of $S_\mathrm{3b}$ are given
in the central part of Table~\ref{tabres1}.

\begin{table}[t!]
\begin{tabular}{|ccccc|ccccc|ccccc|}
\hline
  \multicolumn{15}{|c|}{3$^+_1$ at $-2.28$ MeV (2.22 MeV exc. energy), $S_\mathrm{3b}=-11.5$ MeV }  \\\hline
\multicolumn{5}{|c|}{$^4$H($t,n$)+$\alpha$}    &  \multicolumn{5}{c|}{$^5$He($\alpha,n$)+$t$}  &  \multicolumn{5}{c|}{$^7$Li($\alpha,t$)+$n$}   \\ \hline
 $\ell_x$  &    $j_x$  &  $\ell_y$ & $j_y$ &  weight  &  $\ell_x$  &  $j_x$ &  $\ell_y$ & $j_y$ &  weight    &  $\ell_x$  &  $j_x$ &  $\ell_y$ & $j_y$ &  weight  \\ \hline
    0            &    1    &      2          &      2    &    21.8   & 0  & $\frac{1}{2}$  &  2 & $\frac{5}{2}$  & 16.9   &  1  &  $\frac{3}{2}$    & 1   & $ \frac{3}{2}$  &  70.6 \\
    1            &    1    &      3          &      3    &     7.0     & 1  & $\frac{1}{2}$  & 3 & $\frac{7}{2}$  & 1.1   &  3  &  $\frac{5}{2}$    & 1   &  $\frac{3}{2}$  &  3.2 \\
    1            &    2    &      1          &      1    &    33.0   &  1  & $\frac{3}{2}$  & 1 & $\frac{3}{2}$  &  50.9     & 3  &  $\frac{7}{2}$    & 1   &  $\frac{1}{2}$  &   2.9 \\
    1            &    2    &      3          &      3    &    13.1   &  1  & $\frac{3}{2}$  & 3 & $\frac{5}{2}$  & 2.3   &  3  &  $\frac{7}{2}$    & 1   &  $\frac{3}{2}$  &  20.5 \\
    2            &    2    &      2          &      2    &    1.0     &  1  & $\frac{3}{2}$  &  3 & $\frac{7}{2}$  & 13.6   &    &      &    &    &   \\
    2            &    3    &      0          &      0    &    19.0   &  2  & $\frac{5}{2}$  & 0   &$\frac{1}{2}$  & 11.8       &    &      &   &    &   \\
    3            &    4    &      1          &      1    &    1.9     &   2 &  $\frac{5}{2}$ &  2  & $\frac{5}{2}$  &  1.5      &    &      &   &    &   \\
   \hline
\end{tabular}
\caption{The same as Table~\ref{tab2plus} for the $3^+_1$ resonance 2.28 MeV below the $\alpha$+$t$+$n$ threshold.}
\label{tab31}
\end{table}

In Table~\ref{tab31} we give the contribution of the different components to the $3^+_1$ resonance when $S_\mathrm{3b}$=-11.5 MeV. We see that the contribution from $^7$Li in the $\frac{1}{2}^-$ state disappears, and the wave function is now dominated by $^7$Li in the 
$\frac{3}{2}^-$ and $\frac{7}{2}^-$ states.  In the last two columns of Table~\ref{tabres1}, middle part, we show how these contributions change for different values
of the three-body strength. We observe that the weight of the  $\frac{3}{2}^-$ components  increases when the three-body attraction decreases. Again, this reflects the fact that, for a sufficiently weak three-body attraction, the resonance disappears in the $^7$Li+neutron continuum with $^7$Li in the $\frac{3}{2}^-$ ground state. In this case, the components of $^7$Li in $\frac{1}{2}^-$ and $\frac{5}{2}^-$ states (not given in the table), contribute with only a small percentage to
the resonance wave function.

\begin{figure}[t!]
\centering
\includegraphics[width=\columnwidth]{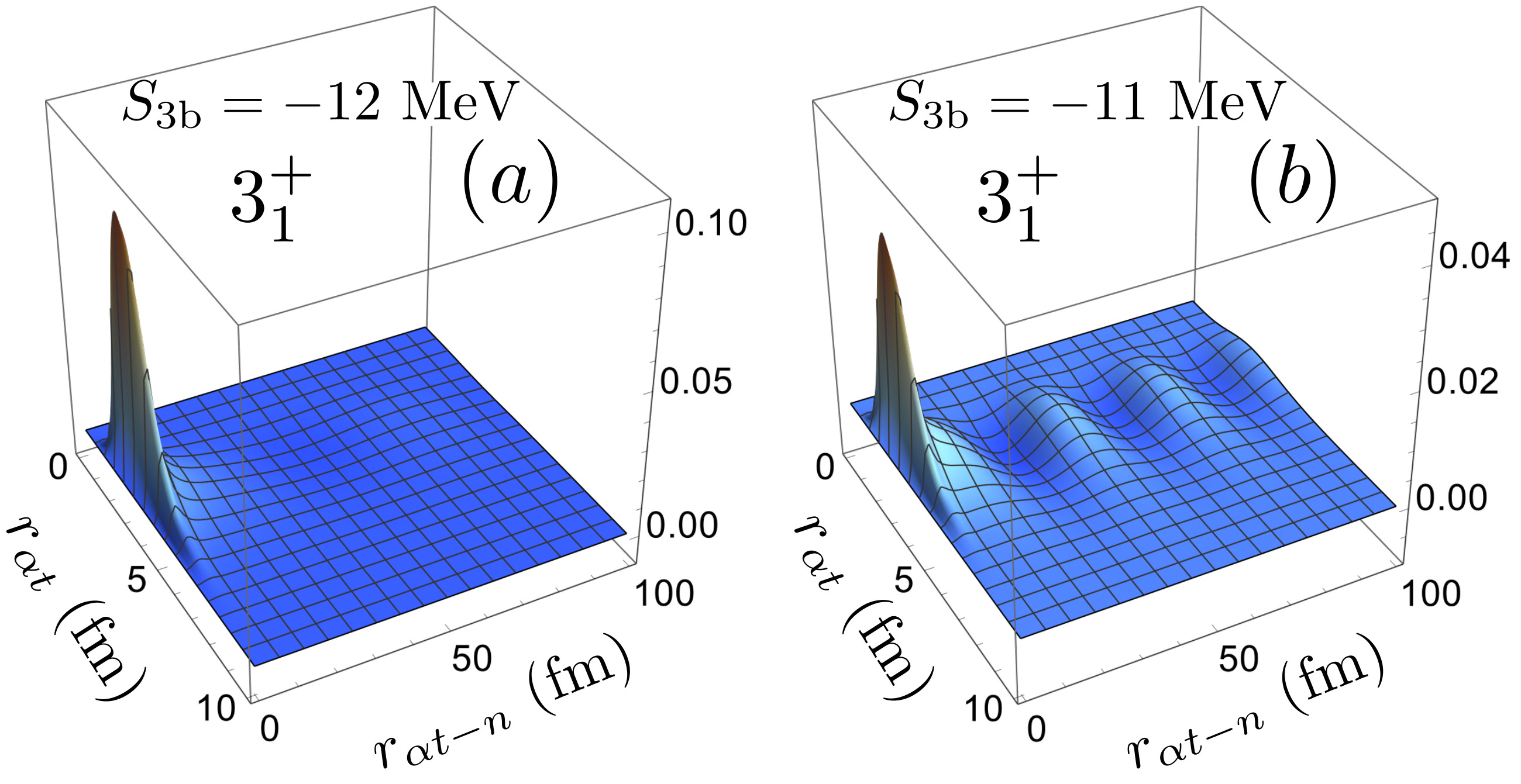}
\caption{The same as Fig.~\ref{fig:3dwf1} but for the $3^+_1$ resonance. The $(a)$ and $(b)$ panels correspond to  
the Gaussian three-body strengths $S_\mathrm{3b}=-12$ MeV and $S_\mathrm{3b}=-11$ MeV, respectively.}
\label{fig:3dwf3}
\end{figure}

For completeness, we show in Fig.~\ref{fig:3dwf3} the same as in Fig.~\ref{fig:3dwf1} but for the $3^+_1$ resonance. The result is very similar to the 1$^+_2$ case. 
For a large three-body 
attraction, $S_\mathrm{3b}=-12$ MeV, Fig.~\ref{fig:3dwf3}a, the wave function is very confined. We get the two root-mean-square radii to be $\langle r_{\alpha t}^2 \rangle^{1/2}=3.4 +i 0.2$ fm
and  $\langle r_{\alpha t-n}^2 \rangle^{1/2}=4.5 + i 3.1$  fm.
When the attraction is progressively reduced to $S_\mathrm{3b}=-11$ MeV,
Fig.~\ref{fig:3dwf3}b, the value of $\langle r_{\alpha
  t}^2 \rangle^{1/2}$ remains very stable, but the neutron is again
more and more delocalized, as shown by the oscillations in the
$r_{\alpha t-n}$ direction of the $F(r_{\alpha t}, r_{\alpha t-n})$
function defined by Eq.(\ref{wf2}).  In contrast to the $1^+_2$
resonance, the barrier for $3^+_1$ is sufficient to maintain the
characteristic behavior of a resonance with a finite width.

\subsubsection{The 0$^+$ states.}

Although there is no experimental evidence  of isospin one 0$^+$ states in the $^8$Li spectrum, several studies \cite{Wiringa2000} pointed out their existence. For this
reason we have also considered this possibility in our calculations. As done for the 1$^+_2$ and 3$^+_1$ states, we show in Fig.~\ref{fig:res1}c the discrete energy spectrum obtained after a complex scaling calculation. In this case the rotation angle is $\theta=0.3$~rads. 

As we can see, the result is very similar to what is found in Fig.~\ref{fig:res1}a for the 1$^+_2$ states. When no three-body interaction is included, we found a slightly bound $0^+_1$
state, barely 120 keV below the $^7$Li+$n$ threshold. Therefore, again, a relatively small three-body repulsion is sufficient to push this 0$^+_1$ state into the resonance region. In particular, the
red dots shown close to the origin in Fig,~\ref{fig:res1}c have been obtained with a three-body strength varying from $S_ \mathrm{3b}=0$ to   $S_ \mathrm{3b}=4$ MeV. In the lower part of Table~\ref{tabres1} we give the resonance
energies and widths for some $S_\mathrm{3b}$ values.  Essentially all
weight of the components correspond to the $\alpha$-$t$ system in
the $\frac{3}{2}^-$ or $\frac{1}{2}^-$ state plus a neutron mostly in
$p$-waves, see Table~\ref{tab0plusa}.

\begin{table}[t!]
\begin{center}
\begin{tabular}{|ccccc|ccccc|ccccc|}
 \hline
 \multicolumn{15}{|c|}{0$^+_1$ at $-1.90$ MeV (2.60 MeV exc. energy, $S_\mathrm{3b}=4.0$ MeV)}  \\\hline
\multicolumn{5}{|c|}{$^4$H($t,n$)+$\alpha$}    &  \multicolumn{5}{c|}{$^5$He($\alpha,n$)+$t$  }  &  \multicolumn{5}{c|}{ $^7$Li($\alpha,t$)+$n$ }   \\    \hline
  $\ell_x$   &  $j_x$  &  $\ell_y$ & $j_y$ &  weight  &  $\ell_x$  &  $j_x$ &  $\ell_y$ & $j_y$ &  weight    &  $\ell_x$  &  $j_x$ &  $\ell_y$ & $j_y$ &  weight  \\ \hline
     1             &   1    &      1          &      1    &    96.7   &  1  & $\frac{1}{2}$  &  1 & $\frac{1}{2}$  & 47.5   &  1  &   $\frac{1}{2}$   & 1   &  $\frac{1}{2}$  &  51.5 \\
     2             &   2    &      2          &      2    &    3.3   &    1 &  $\frac{3}{2}$ &   1 & $\frac{3}{2}$  & 50.6   &  1  & $\frac{3}{2}$  &   1    &  $\frac{3}{2}$ &  46.6          \\
                   &        &                 &           &         &  2 &  $\frac{3}{2}$ &   2 & $\frac{3}{2}$  &  1.8  &     &   &     &   &           \\
     \hline
  \multicolumn{15}{c}{ }  \\ \hline
\multicolumn{15}{|c|}{0$^+_2$ at 1.10 MeV (5.60 MeV exc. energy)}  \\\hline
\multicolumn{5}{|c|}{$^4$H($t,n$)+$\alpha$}    &  \multicolumn{5}{c|}{$^5$He($\alpha,n$)+$t$  }  &  \multicolumn{5}{c|}{ $^7$Li($\alpha,t$)+$n$ }   \\  \hline
  $\ell_x$   &  $j_x$  &  $\ell_y$ & $j_y$ &  weight  &  $\ell_x$  &  $j_x$ &  $\ell_y$ & $j_y$ &  weight    &  $\ell_x$  &  $j_x$ &  $\ell_y$ & $j_y$ &  weight  \\ \hline
     0             &   0    &      0          &      0    &    3.6   &     1  & $\frac{3}{2}$  &  1 & $\frac{3}{2}$  & 99.9   &  0  &   $\frac{1}{2}$   & 0   &  $\frac{1}{2}$  &  36.7 \\
     1             &   1    &      1          &      1    &    44.8   &    &    &    &   &    &  1  & $\frac{1}{2}$  &   1    &  $\frac{1}{2}$ &  6.8          \\
      2            &   2    &      2          &      2    &    32.0   &    &    &    &   &   &   1  & $\frac{3}{2}$  &   1  & $\frac{3}{2}$  &  11.6         \\
      3            &   3    &      3          &      3    &    19.6   &    &   &    &   &    &   2 &  $\frac{3}{2}$    &   2  &  $\frac{3}{2}$ &     6.8      \\
                   &        &                 &            &               &    &   &    &   &    &   2 &  $\frac{5}{2}$    &   2  &  $\frac{5}{2}$ &     23.7      \\
                   &        &                 &            &               &    &   &    &   &    &   3 &  $\frac{5}{2}$    &   3  &  $\frac{5}{2}$ &     4.1     \\
                  &        &                 &            &               &    &   &    &   &    &   3 &  $\frac{7}{2} $   &   3  &  $\frac{7}{2}$ &     10.3     \\
   \hline
\end{tabular}
\end{center}
\caption{The same as Table~\ref{tab2plus} for the $0^+_1$ rand $0^+_2$ resonances at 2.07 MeV below and 1.10 MeV above the $\alpha$+$t$+$n$ threshold.}
\label{tab0plusa}
\end{table}

In the upper part of Table~\ref{tab0plusa} we give the contributions of the components entering in the $0^+_1$ states, when located at 1.90 MeV below the 
$\alpha$+$t$+$n$ threshold. The result is not very different from the one obtained for the $1^+_2$ state shown in Table~\ref{tab1plusa}, i.e., in the first Jacobi set an
almost purely $^4$H and an $\alpha$ in a $p$-wave. Furthermore,
in the second and third Jacobi sets, the wave function is dominated by $^5$He in the
$\frac{1}{2}^-$ and $\frac{3}{2}^-$ states and $^7$Li in the $\frac{1}{2}^-$ or $\frac{3}{2}^-$
states, respectively, while the third particle is in continuum
$p$-waves.

As observed in the $1_2^+$ case, an increase of the
three-body repulsion implies as well an increase of the contribution coming from the components having the $\alpha$-$t$ system in the 
$\frac{1}{2}^-$ state, and a corresponding decrease of the $\frac{3}{2}^-$ components.  In fact, the $0_1^+$ resonance tends, as the $1^+$ case, to be approaching the cut in the energy
plane associated to the continuum states of $^7$Li in the $\frac{1}{2}^-$ state plus a neutron.  It also disappears there, since the $0^+_1$ resonance is not found
when for $S_\mathrm{3b} > 4$~MeV.

Therefore, in our calculations a $0^+$ resonance appears with energy and width 
 very similar to the one obtained for the $1^+_2$ state (although experimentally the
1$^+_2$ state is a few hundreds of keV higher in energy than the computed one). 
Even the three-body wave
function of the $0^+_1$ state is also similar the one of the 1$^+_2$
state. The representation of the $F(r_{\alpha t}, r_{\alpha t-n})$
function defined in Eq.(\ref{wf2}) is hardly distinguishable by naked
eye from the one shown in Fig.~\ref{fig:3dwf1} for the $1^+_2$ state.

\begin{figure}[t!]
\includegraphics[width=4.5cm]{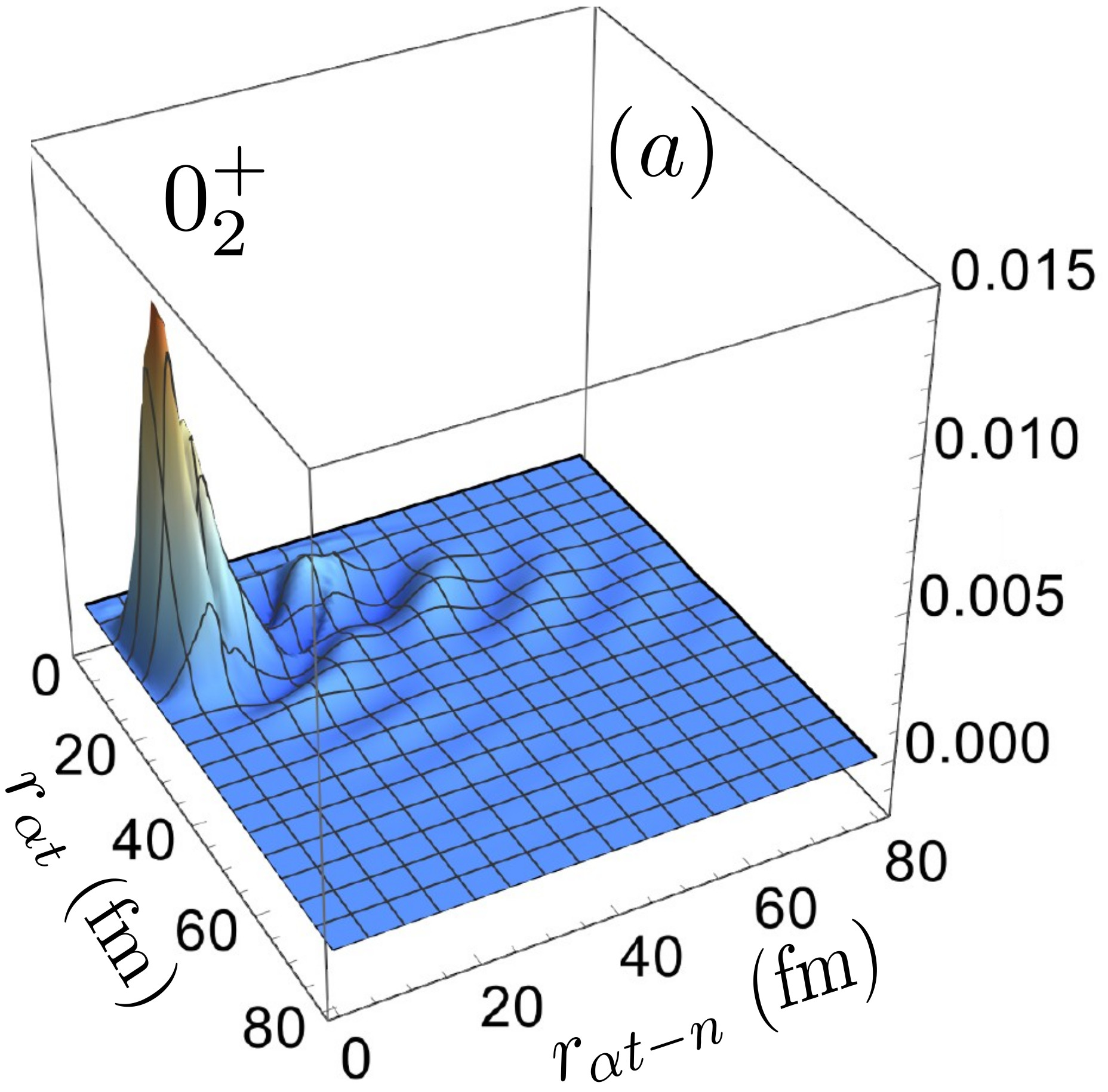}
\includegraphics[width=4cm]{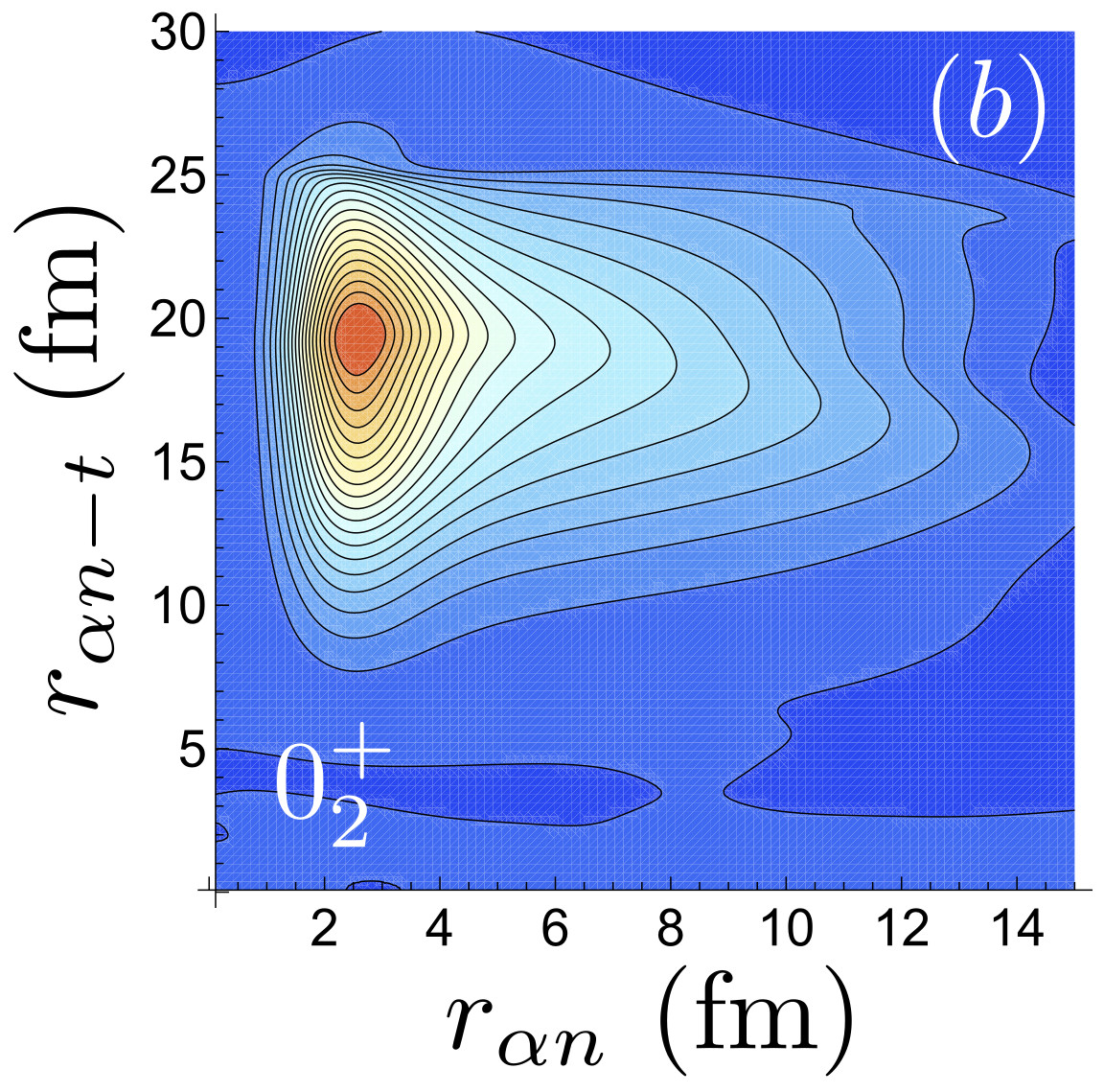}
\caption{$(a)$: The same as Fig.~\ref{fig:3dwf1} but for the  $0^+_2$ resonance. The result is independent of the  $S_\mathrm{3b}$ value.
$(b)$: Contour plot of the probability function in Eq.(\ref{wf2}), but in the Jacobi set where $\bm{x}$ connects the $\alpha$ and the neutron.}
\label{fig:3dwf02}
\end{figure}

It is important to note that we have found a second $0^+_2$ resonance
with energy and width $(E_R,\Gamma_R)=(1.10,0.96)$ MeV, see
Fig.~\ref{fig:res1}c.  The position of this resonance is pretty much
independent of the three-body strength, $S_\mathrm{3b}$. The
contribution of the different components for this resonance is shown
in the lower part of Table~\ref{tab0plusa}. From the table, we can
immediately see that this resonance almost fully corresponds to a
structure with $^5$He in the $\frac{3}{2}^-$ state and the triton in continuum
$p$-waves. In fact, since the energy of the $\frac{3}{2}^-$ resonance in $^5$He
is only 0.80 MeV above the threshold, we have that the computed energy of the $0^+_2$
state is only 0.3 MeV above the threshold corresponding to
$^5$He$(3/2^-)$ plus a triton.  Not surprisingly, the first and third Jacobi sets
in Table~\ref{tab0plusa} exhibit many components.

This is also seen 
in the $F(r_{\alpha t}, r_{\alpha t-n})$ function, Eq.(\ref{wf2}), which for this resonance is shown in Fig.~\ref{fig:3dwf02}a. As we can see, the wave function
is now pretty much extended in both directions, $r_{\alpha t} $ and $r_{\alpha t-n}$, and contrary to what is seen in Fig.~\ref{fig:3dwf1},  the oscillation of the
wave function appears as well in the $\alpha$-$t$ relative distance, which reflects the lack of $\alpha$-triton binding. The structure of the $0^+_2$ resonance
is however better seen when looking at the the probability function (\ref{wf2}) but in the second Jacobi set, where the relative coordinates $r_{\alpha n}$ and $r_{\alpha n-t}$
are used. The contour plot of this probability function is shown in Fig.~\ref{fig:3dwf02}b. We see that its maximum it located at a small $\alpha$-$n$ distance,
$r_{\alpha n} \approx 2.5$ fm, but  at a large relative distance between the triton and the $\alpha$-$n$ center of mass. Due to the resonant character of the
$\frac{3}{2}~^-$ state in $^5$He, the probability function extends as well in the $r_{\alpha n}$ coordinate.

\section{$^8$Li as an $\alpha$+$d$+$^2n$ system}
\label{sec:ad2n}

As shown in the previous section, the $\alpha$+$t$+$n$ structure does not provide information about
$^8$Li states above the three-body breakup threshold. This leads us to
investigate those higher-lying states as described with a different
three-body clusterization.  In particular, we shall consider the
$\alpha$+$d$+$^2n$ structure using the two-body interactions described
in Section~\ref{secad2n}.  Other possible three-body divisions could
be $^6$Li+$n$+$n$ or $^6$He+$n$+$p$. However, combining two neutrons into
a dineutron and a neutron and proton into the deuteron, allow
inclusion of excited states of the subsystems $^6$Li($\alpha+d$) or
$^6$He($\alpha+^2n$) in the same three-body calculations.  In fact,
the chosen clusterization ($\alpha$+$d$+$^2n$) does not seem to be a
substantial limitation, since both $d$ and $^2n$ can be considered to
act as effective particles, where their intrinsic structures are
unimportant.

We note, see Fig.~\ref{fig:sch}, that all isospin $1$ states except
$1^+_4$, are below the $^6$Li+$^2n$ threshold.  (The highest $0^+$
state has isospin 2, and is beyond the scope of the present three-body
work).  Therefore, all the resonances in Fig.~\ref{fig:sch}, except
the $1^+_4$ state, are not allowed to decay into their constituents
$^6$Li, $^6$He, $d$ and $^2n$.  Thus, the interesting states are bound
with respect to the three-body configuration, $\alpha$+$d$+$^2n$.
Only the $1^+_4$ resonance can decay through the $^6$Li+$^2n$ channel.

\subsection{The 1$^+$ states}

The three-body calculation gives rise to two 1$^+$ states. When no three-body
force is used, we get them with separation energies $-4.2$ MeV and
$-1.6$ MeV below the $\alpha$+$d$+$^2n$ threshold. These two energies
are not very far from the experimental energies of the $1^+_3$ and
$1^+_4$ states seen above the $\alpha$+$t$+$n$ threshold, which are
$-5.36$ MeV and $-1.09$ MeV, respectively.  Therefore, a little
three-body attraction or repulsion are sufficient to fit the
experimental data with the calculations.  To be precise,
$S_\mathrm{3b}=-2.9$ MeV and $S_\mathrm{3b}= +2.5$ MeV do the job for
each of the two cases.

As mentioned above, the second of these $1^+$ states, $1^+_4$, is actually a resonance in the  $\alpha$+$d$+$^2n$ configuration, which is only 0.38 MeV above the
$^6$Li+$^2n$ threshold. The computed width for this channel decay is  pretty small, of only about 10 keV, much smaller than the experimental width of about
1 MeV given in Ref.~\cite{Tilley2004}.

\begin{table}[t!]
\begin{center}
  \begin{tabular}{|ccccc|ccccc|ccccc|}
    \hline 
\multicolumn{15}{|c|}{1$^+_3$ at $-5.36$ MeV (5.40 MeV exc. energy), $S_\mathrm{3b}=-2.9$ MeV }  \\\hline
\multicolumn{5}{|c|}{$^4$H($d$,$^2n$)+$\alpha$ }    &  \multicolumn{5}{c|}{$^6$Li($\alpha,d$)+$^2n$ }  &  \multicolumn{5}{c|}{$^6$He($\alpha$,$^2n$)+$d$ }  \\  \hline
  $\ell_x$   &  $j_x$  &  $\ell_y$ & $j_y$ &  weight  &  $\ell_x$  &  $j_x$ &  $\ell_y$ & $j_y$ &  weight    &  $\ell_x$  &  $j_x$ &  $\ell_y$ & $j_y$ &  weight  \\ \hline
     0             &   1    &      0          &      0    &    97.5   & 0  & 1  &  0 & 0  & 98.3   &  0  &  0    & 0   &  1  &  99.5 \\
   \hline
  \multicolumn{15}{c}{ }  \\
   \hline
\multicolumn{15}{|c|}{1$^+_4$ at $-1.09$ MeV (9.67 MeV exc. energy), $S_\mathrm{3b}=2.5$ MeV}  \\\hline
\multicolumn{5}{|c|}{$^4$H($d$,$^2n$)+$\alpha$ }    &  \multicolumn{5}{c|}{$^6$Li($\alpha,d$)+$^2n$ }  &  \multicolumn{5}{c|}{$^6$He($\alpha$,$^2n$)+$d$ }  \\  \hline
  $\ell_x$  &  $j_x$  &  $\ell_y$ & $j_y$ &  weight  &  $\ell_x$  &  $j_x$ &  $\ell_y$ & $j_y$ &  weight    &  $\ell_x$  &  $j_x$  &  $\ell_y$ & $j_y$ &  weight  \\ \hline
     0            &     1    &      2          &      2    &    37.3   & 0  & 1  &  2 &  2  & 74.1   &  0  &  0    & 2   &  1  &  2.8 \\    
     1            &     0    &      1          &      1    &    15.7   & 1  & 0  &  1 &  1  &  6.2    &  1  &  1    & 1   &  0  &  1.1 \\
     1            &     1   &       1          &      1    &    11.8   & 1  & 1  &  1 &  1  &  4.7    &  2  &  2    & 0   &  1  &  93.5 \\
     1            &     2   &       3          &      3    &      3.9   & 1  & 2  &  3 &  3  &  6.7    &       &        &      &      &       \\
     2            &     1   &       0          &      0    &     23.2  &2  & 1  &  0 &  0  &  5.7    &       &        &      &      &       \\
     2            &     2   &       2          &      2    &      2.3  &    &    &    &     &      &       &        &      &      &       \\
    3            &      2   &       1          &      1    &      2.3  &    &    &    &     &      &       &        &      &      &       \\
 \hline
\end{tabular}
\end{center}
\caption{For each of the  three Jacobi sets, the upper and lower parts give, respectively, the components contributing with more than 1\% to the $1^+_3$  and $1^+_4$
states at 5.36 MeV  and 1.09 MeV below the $\alpha$+$d$+$^2n$ threshold in $^8$Li, respectively.
The first line in each of the tables gives the computed separation energy referred to the three-body threshold, the corresponding excitation
energy, and the strength, $S_\mathrm{3b}$, used in three-body potential in Eq.(\ref{3bdf}).}
\label{tab1plusb}
\end{table}

In table~\ref{tab1plusb} we give the components that contribute to the wave function of both these 1$^+$ states with more than 1\%. As we can see in the upper part, for the
state at $-5.36$ MeV one single component on each of the three Jacobi sets give most of the wave function.  They are characterized by relative $s$-waves between 
the particles, therefore the total orbital angular momentum is 0, and, necessarily, the total spin is equal to 1.

For the 1$^+_4$ state, shown in the lower part of the table, the contribution is divided
into several components, especially the first Jacobi set of
$^4$H$+\alpha$.  In the second Jacobi set the main component is $^6$Li in
$1^+$ surrounded by $^2n$ in $d$-waves.  From the low-right part of
table~\ref{tab1plusb} we see that most of the $1^+_4$ wave function
corresponds to $^6$He in the $2^+$ state and the deuteron in a
relative $s$-wave. 
In other words, this resonance is by far dominated
by  components  with total orbital angular
momentum equal to 2 and intrinsic total
spin equal to 1.

\subsection{The 2$^+$ and 3$^+$ states}

When no three-body force is included, the calculation gives rise to a
3$^+$ state at 2.87~MeV below the $\alpha$+$d$+$^2n$ threshold.  This
state is not far from the state at $3.66$~MeV below the
$\alpha$+$d$+$^2n$ threshold, where no measured angular momentum and
parity are given, see Fig.~\ref{fig:sch}.  A three-body attraction of
$- 2.15$ MeV puts the computed state at the given energy.  Instead, to
locate this state at 4.66 MeV below the threshold, which is the
experimental state whose tentative assigned angular momentum is 3, we
need a three-body strength of $-4.7$ MeV.

\begin{table}[t!]
\begin{center}
  \begin{tabular}{|ccccc|ccccc|ccccc|}
     \hline
\multicolumn{15}{|c|}{3$^+_2$ at $-4.66$ MeV (6.10 MeV exc. energy), $S_\mathrm{3b}=-4.7$ MeV}  \\\hline
\multicolumn{5}{|c|}{$^4$H($d$,$^2n$)+$\alpha$ }    &  \multicolumn{5}{c|}{$^6$Li($\alpha,d$)+$^2n$ }  &  \multicolumn{5}{c|}{$^6$He($\alpha$,$^2n$)+$d$ }   \\  \hline
  $\ell_x$  &    $j_x$  &  $\ell_y$ & $j_y$ &  weight  &  $\ell_x$  &  $j_x$ &  $\ell_y$ & $j_y$ &  weight    &  $\ell_x$  &  $j_x$ &  $\ell_y$ & $j_y$ &  weight  \\ \hline
     0            &    1    &      2          &      2    &    73.1   & 0  & 1  &  2 & 2  & 32.0   &  0  &  0    & 2   &  3  &  39.4 \\
     2            &    3    &      0          &      0    &    22.6   &  1  & 2  &  1 & 1  &  6.2     &  1  &  1    & 1   &  2  &   8.5 \\
     2            &    3    &      2          &      2    &    2.3   & 2  & 3  &  0 & 0  & 59.5   &  2  &  2    & 0   &  1  &  49.7 \\
   \hline
  \multicolumn{15}{c}{ }  \\  \hline
\multicolumn{15}{|c|}{2$^+_2$ at $-3.71$ MeV (7.05 MeV exc. energy), $S_\mathrm{3b}=-4.7$ MeV}  \\\hline
\multicolumn{5}{|c|}{$^4$H($d$,$^2n$)+$\alpha$ }    &  \multicolumn{5}{c|}{$^6$Li($\alpha,d$)+$^2n$ }  &  \multicolumn{5}{c|}{$^6$He($\alpha$,$^2n$)+$d$ }   \\  \hline
  $\ell_x$  &   $j_x$  &  $\ell_y$ & $j_y$ &  weight  &  $\ell_x$  &  $j_x$ &  $\ell_y$ & $j_y$ &  weight    &  $\ell_x$  &  $j_x$  &  $\ell_y$ & $j_y$ &  weight  \\ \hline
     0            &   1    &      2          &      2    &    60.9   & 0  & 1  &  2 &  2  & 62.0   &  0  &  0    & 2   &  2  &  10.9 \\    
     1            &   1    &      1          &      1    &     9.4    & 1  & 1  &  1 &  1  &  9.9     &  1  &  1    & 1   &  1  &  1.3 \\
     1            &   2   &       1          &      1    &    3.1     & 1  & 2  &  1 &  1  &  3.3     &  2  &  2    & 0   &  1  &  86.5 \\
     2            &   1   &       2          &      2    &     1.3    & 1  & 2  &  3 &  3  &  1.9    &       &        &      &      &       \\
     2            &   2   &       0          &      0    &     20.5 & 2  & 2  &  0 &  0  &  20.3    &       &        &      &      &       \\
     2            &   3   &       2          &      2    &      1.5  &    &    &    &     &      &       &        &      &      &       \\
 \hline
\end{tabular}
\end{center}
\caption{The same as Table~\ref{tab1plusb} for the $3^+_2$ and
$2^+_2$ states at 4.66 MeV and 3.62MeV below the $\alpha$+$d$+$^2n$ threshold in $^8$Li.}
\label{tab3plusb}
\end{table}

In the upper part of Table~\ref{tab3plusb} we give the weights of the
different components entering in the wave function with more than 1\%
of the contribution.  We see
that, as in the case of the $1^+_4$ state, most of the wave function
($\sim 90\%$) results from the coupling of total orbital angular
momentum equal to 2, either $(2,0)$ or $(0,2)$, and the deuteron spin equal
to 1 (both the $\alpha$-particle  and the $^2n$ have zero spin).

The fact that the $1^+_4$ and $3^+_2$ states are mainly produced by the coupling of the orbital angular momentum 2 and the total spin 1, immediately suggests
that there should also be a $2^+$ state coming from the same coupling. Furthermore, this state should very likely be located between the $3^+_2$ and $1^+_4$ states.  
 In fact our calculation predicts such a $2^+$ state, $2^+_2$, at 2.12 MeV below the three-body $\alpha$+$d$+$^2n$ threshold, and
 at $-3.71$ MeV, if we use the same three-body force that locates the $3_2^+$ state at $-4.66$ MeV.

The contribution from the different components in the wave function of
the $2_2^+$ state is given in the lower part of Table~\ref{tab3plusb}.
We see that, as expected, this state is also
arising from the coupling of total orbital angular momentum equal to
2, and total spin equal to 1.   Experimentally, see
Fig.~\ref{fig:sch}, $^8$Li is known to have two states with unassigned
spin and parity 1.76 MeV and 3.66 MeV below the $\alpha$+$d$+$^2n$
threshold, respectively. According to our calculation, the one we have obtained could correspond
to the state at $-3.66$ MeV.

\subsection{The 4$^+$ and 0$^+$ states}

\begin{table}[t!]
\begin{center}
  \begin{tabular}{|ccccc|ccccc|ccccc|}
     \hline
\multicolumn{15}{|c|}{4$^+_1$ at $-4.23$ MeV (6.53 MeV exc. energy), $S_\mathrm{3b}=-18.2$ MeV}  \\\hline
\multicolumn{5}{|c|}{$^4$H($d$,$^2n$)+$\alpha$ }    &  \multicolumn{5}{c|}{$^6$Li($\alpha,d$)+$^2n$ }  &  \multicolumn{5}{c|}{$^6$He($\alpha$,$^2n$)+$d$ }   \\  \hline
  $\ell_x$   &  $j_x$  &  $\ell_y$ & $j_y$ &  weight  &  $\ell_x$  &  $j_x$ &  $\ell_y$ & $j_y$ &  weight    &  $\ell_x$  &  $j_x$ &  $\ell_y$ & $j_y$ &  weight  \\ \hline
     1             &   1    &      3          &      3    &    25.3   &  1  & 2  &  3 & 3  &  1.7   &    1  &  1    & 3   &  4  &    1.6 \\
     1             &   2    &      3          &      3    &    40.6   &  2  & 2  &  2 & 2  &  7.1     &  2  &  2    & 2   &  2  &   7.9 \\
     3             &   3    &      1          &      1    &     2.3   &   2  & 3  &  2 & 2  & 75.7   &   2  &  2    & 2   &  3  &  76.3 \\  
     3             &   4    &      1          &      1    &    30.8   & 3  & 4  &  1 & 1  & 13.3   &    3  &  3   & 1   &  1  &  2.4 \\
                    &         &                    &            &                &     &      &    &     &              &    3  &  3   & 1   &  2  &  9.9 \\
                    &         &                    &            &                &     &      &    &     &              &     3  &  3   & 3  &   4  &  1.4 \\
   \hline
  \multicolumn{15}{c}{ }  \\  \hline
\multicolumn{15}{|c|}{0$^+_3$ at $-1.76$ MeV (9.00 MeV exc. energy), $S_\mathrm{3b}=-17.6$ MeV}  \\\hline
\multicolumn{5}{|c|}{$^4$H($d$,$^2n$)+$\alpha$ }    &  \multicolumn{5}{c|}{$^6$Li($\alpha,d$)+$^2n$ }  &  \multicolumn{5}{c|}{$^6$He($\alpha$,$^2n$)+$d$ } \\  \hline
  $\ell_x$   &  $j_x$  &  $\ell_y$ & $j_y$ &  weight  &  $\ell_x$  &  $j_x$ &  $\ell_y$ & $j_y$ &  weight    &  $\ell_x$  &  $j_x$  &  $\ell_y$ & $j_y$ &  weight  \\ \hline
     1             &   1    &      1          &      1    &    95.9   & 1  & 1  &  1 &  1  & 15.0     &  1  &  1    & 1   &  1  &  10.1 \\    
     3             &   3    &      3          &      3    &    3.7    & 2  & 2  &  2 &  2  &  81.3    &  2  &  2    & 2   &  2  &  86.2 \\
                     &         &                   &             &               & 3  & 3  &  3 &  3  &  3.7       &  3  &  3    & 3   &  3  &  3.7\\
 \hline
\end{tabular}
\end{center}
\caption{The same as Table~\ref{tab1plusb} for  the $4^+_1$ and
$0^+_3$ states at 4.23 MeV below and 2.66 MeV above the $\alpha$+$d$+$^2n$ threshold in $^8$Li.}
\label{tab4plusb}
\end{table}

Finally, we have also investigated the possibility of existence of
4$^+$ and $0^+$ states within the $\alpha$+$d$+$^2n$ structure.
Without any three-body force, a narrow $4^+_1$ resonance has been
found at 1.18 MeV above the $\alpha$+$d$+$^2n$ threshold, and only 60
keV wide.  This energy exceeds the experimental value, which is 4.23
MeV below the three-body threshold, see Fig.~\ref{fig:sch}. The
consequence is that a large three-body attraction, $S_\mathrm
{3b}=-18.2$ MeV, is required in order to fit the experimental energy.

In any case, the computed energy for the $4^+_1$ state has the
additional uncertainty coming from the fact that higher partial waves
between the particles are important, but only the components with
$\ell \leq 3$ have been included.  In fact, as we can see in the upper
part of Table~\ref{tab4plusb}, the $\ell_x=3$ components contribute
with about $30\%$ of the norm in the first Jacobi set, and about
$13\%$ in the other two.  These $\ell=3$ two-body potentials have not
been adjusted to any $^4$H, $^6$Li, or $^6$He experimental data, but
they have simply been taken equal to the one for $\ell=2$.

A $0^+_3$ resonance has also been found at 2.66~MeV above the $\alpha$+$d$+$^2n$ threshold with a width of 0.46 MeV (with no three-body force). 
Since the $\alpha$+$d$+$^2n$ structure is only consistent with total isospin 1, this state can not correspond to the only $0^+$ state known experimentally,
 which has isospin 2. In principle, this state could correspond to one of the $^8$Li states, known experimentally, but without assigned angular momentum. The closest of these
states to the computed $0^+_3$ resonance is the one 1.76 MeV below the $\alpha$+$d$+$^2n$.  As in the $4^+_1$ case, a large three-body force 
strength, $S_\mathrm{3b}=-17.6$ MeV is needed to reach this energy. For completeness we give in the lower part of Table~\ref{tab4plusb} the contribution of the different components to this state.

\section{Structure of calculated states}
\label{sec:comp}

Two structures based on different three-body clusters are investigated
in details. The angular momenta and parities of the resulting states
are surveyed in the next subsections. Afterwards, we compare with
previous work providing the low-energy spectrum of $^8$Li.

\subsection{The $\alpha$+$t$+$n$ spectrum}

\begin{figure}[t!]
\centering
\includegraphics[width=8cm]{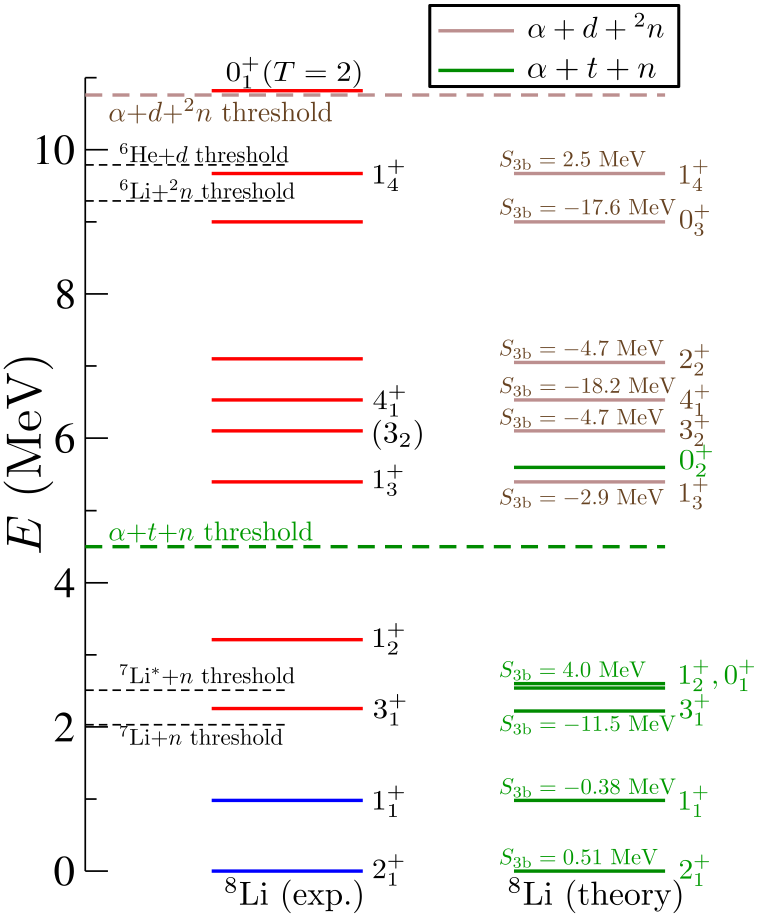}
\caption{Comparison between the experimental spectrum in $^8$Li \cite{Tilley2004}  (left part) and the one obtained in the numerical calculations (right part). The
green and brown bars represent the computed states obtained assuming an $\alpha$+$t$+$n$ and  an $\alpha$+$d$+$^2n$ three-body structure, respectively.
For each of the states we give the value of the three-body potential strength, $S_\mathrm{3b}$, used to fit the energy (the $0^+_2$ state is to a large extent
independent of  $S_\mathrm{3b}$).}
\label{fig:sch3}
\end{figure}

The $^8$Li states obtained after the $\alpha$+$t$+$n$ configuration are represented by the green bars in the right part of Fig.~\ref{fig:sch3}.  The main conclusion 
is that such a three-body structure describes reasonably well the experimental spectrum (right part of the figure) below the $\alpha$+$t$+$n$ threshold. 

The $2^+_1$ and $1^+_1$ bound states are nicely reproduced, and a mild
three-body force is sufficient to fit the experimental energies and
known ground state size. The measured surprisingly strong
$B(E2,2^+\rightarrow 1^+)$ value exceeds the calculations
substantially.  Previous calculations of the same transition strength \cite{Pastore2013,Caprio2022}
present the same discrepancy.

For the resonant states a somewhat stronger three-body force is required.
When using it, the experimental energy and width of the  $3^+_1$ resonance
are reasonably well reproduced. However, the $1^+_2$ resonant state
disappears into the continuum before the experimental value is reached.
The consequence is that, even if the resonance can be located above
the $^7$Li*($1/2^-$)+$n$ threshold,
the energy is about 0.35 MeV smaller than the experimental value.

Together with the experimentally known resonances, the method has
predicted two $0^+$ resonances. The first one has a behavior very similar to the  $1^+_2$ state.
Similar three-body potentials  produce similar three-body energies and widths
for both states. The $0^+_2$ resonance
has been obtained about 1 MeV above the $\alpha$+$t$+$n$ threshold,
with a width of almost $1$~MeV. This is the only state above the three-body
threshold consistent with the $\alpha$+$t$+$n$ structure.

It is also important to mention that, although we have looked for them, neither $2^+$ nor $4^+$ resonances have been obtained with the $\alpha$+$t$+$n$ picture.

\subsection{The $\alpha$+$d$+$^2n$ spectrum}

All the $^8$Li states obtained assuming an $\alpha$+$d$+$^2n$
configuration are represented in the right part of Fig.~\ref{fig:sch3}
by the brown bars.  As we can see, all these states are above the
$\alpha$+$t$+$n$ threshold, and after tuning the energy by means of
the three-body potential, all the experimental states, $1^+_3$,
$(3_2)$, $4^+_1$, and $1^+_4$, can be reproduced. The $4^+_1$ state is
computed with limited partial wave components, i.e. with only $\ell_x,
\ell_y \le 3$, since higher two-body angular momenta are not known,
and expected to provide only moderate contributes.

Together with these states, we have also found $2_2^+$ and $0^+_3$
states.  The energy of the $2_2^+$ state must lie between the $3^+_2$
and $1^+_4$ energies, and it could easily correspond to one of the two
$^8$Li states known experimentally, but without specified spin and
parity quantum numbers. The $0_3^+$ state, which is the third computed $0^+$ state, has been
obtained above the $\alpha$+$d$+$^2n$ threshold, when no three-body
interaction is used.  It is about 4 MeV too high, but it could 
correspond to the second of the experimental states without assigned
spin and parity.

\subsection{Comparison with previous works}

To our knowledge, the only work were the full spectrum of $^8$Li has been systematically
investigated is the one presented in Ref.~\cite{Wiringa2000}. In this work they report on
quantum Monte Carlo calculations using the Argonne $v_{18}$ nucleon-nucleon 
and Urbana IX three-nucleon potentials.

\begin{figure}[t!]
\includegraphics[width=7.5cm]{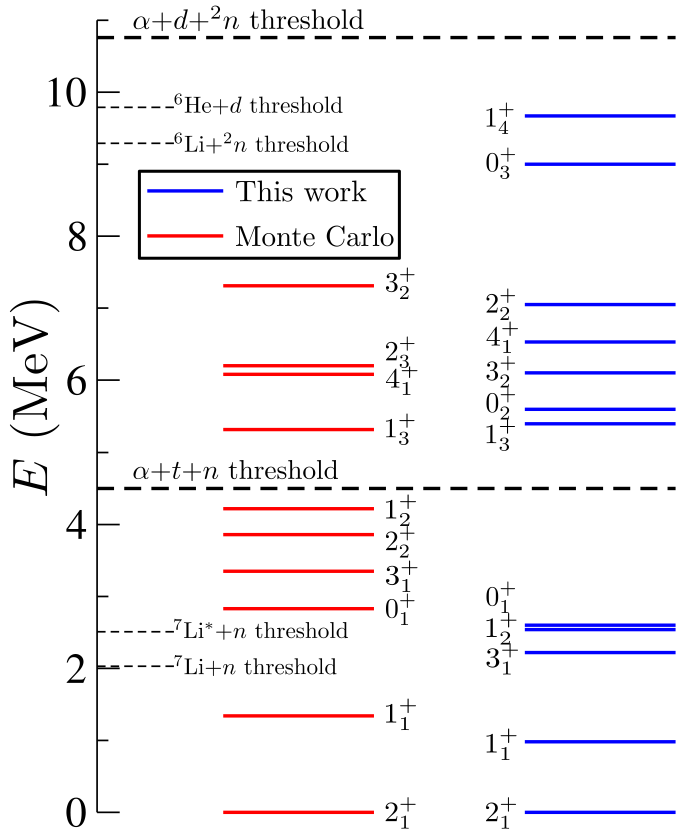}
\caption{Comparison between the computed $^8$Li spectrum obtained in this work (right) and the one obtained after 
the quantum Monte Carlo calculation shown in Ref.~\cite{Wiringa2000} (left).}
\label{fig:comp}
\end{figure}

In Fig.~\ref{fig:comp} we compare the spectrum obtained in this work (right) and the one
given in Ref.~\cite{Wiringa2000} (left). To compare them we should not focus on the specific energy values, 
since in this work we have made use of the freedom provided by the three-body potential in order to,
when possible, fit the experimental energies. We should then focus on the general features of both spectra.

If we first compare just the angular momenta of the computed states, we can see in Fig.~\ref{fig:comp} that 
the two main differences appear in the $2^+$ and $0^+$ excited states. 
For the $2^+$ states the quantum Monte Carlo calculation predicts a resonance below the $\alpha$+$t$+$n$ threshold
that is not present after the three-body calculations presented here. Conversely, the three-body calculation gives rise
to two $0^+$ resonances above the  $\alpha$+$t$+$n$ threshold that are not found in Ref.~\cite{Wiringa2000}.
For the other states, $1^+$ and $4^+$, both calculations are in reasonable agreement (except for the fact that a
fourth $1^+$ state is found in our calculation).

Another difference is that the quantum Monte Carlo calculation does not predict any state close to the
$\alpha$+$d$+$^2n$ threshold. In fact, only four states are predicted above the $\alpha$+$t$+$n$ threshold,
three states less than in our three-body calculation. 

The reason for this discrepancy could be the fact that,
as suggested from our calculations, the states above the $\alpha$+$t$+$n$ threshold correspond to an
$\alpha$+$d$+$^2n$ configuration (except the $0^+_2$ state). In our calculations this configuration is imposed by the procedure itself,
whereas in Ref.~\cite{Wiringa2000} the four nucleons out of the $\alpha$-particle can combine in different
ways, allowing the mixing of different cluster structures. However, when this is done, basically all the states
obtained in \cite{Wiringa2000} are by far dominated by what they call the $[31]$ cluster structure of the
four nucleons, which can be interpreted as a triton+neutron configuration. In principle the $[22]$ configuration,
interpreted as deuteron-dineutron configuration, is also allowed, but the state containing the largest contribution
of this configuration is the $3^+_2$ resonance, which is the highest state that they obtain, and which 
contains only about $8\%$ of such a configuration. 

It is then evident that the states obtained above the $\alpha$+$t$+$n$ threshold have a very different structure in 
both calculations. Although in this work we have constructed the states with a given three-body configuration, without
the possibility of mixing them, the reasonable agreement with the experimental data that we have obtained (Fig.~\ref{fig:sch3})
could be an indication that the calculations shown 
in Ref.~\cite{Wiringa2000} underestimate the contribution of the deuteron-dineutron cluster configuration.

\section{Summary and conclusions}
\label{sec:sum}

We have investigated the $^8$Li spectrum assuming two possible
three-body configurations, namely, $\alpha$+$t$+$n$ and
$\alpha$+$d$+$^2n$. With these configurations we restrict ourselves to
isospin 1 states.  The numerical method is the adiabatic
hyperspherical expansion for bound states and for resonances  we use 
the same  method but combined with complex scaling.

The $\alpha$+$t$+$n$ configuration describes rather well  the four
states below the breakup threshold in the known experimental spectrum.
In addition, the calculation also provided two $0^+$ states. The
second of these states, $0_2^+$, is the only state obtained above the
$\alpha$+$t$+$n$ threshold.  The bound $^8$Li states, below the lowest
break threshold into $^7$Li and a neutron, are well reproduced and a
small three-body force is sufficient to reproduce the experimental
energies.

The lowest computed $1_2^+$ and $0_1^+$ resonances require a small three-body repulsion to show up.
 From the calculation, both these resonances have similar energies and widths.
 However, the measured energy of the $1_2^+$ state is $0.7$~MeV above the 
threshold of $^7$Li*$(1/2^-)$ plus a neutron, and this energy 
 can not be reached in the calculation by use of the three-body potential, since both these states, 
$1_2^+$ and $0_1^+$, disappear into the continuum when the threshold is
approached. The computed $1_2^+$ state is at most 0.35 MeV above
the $^7$Li*$(1/2^-)$+$n$ threshold.

Contrary to the $1_2^+$ and the $0_1^+$ resonances, the experimental energy and
width of the $3^+_1$ resonance are rather well reproduced in the calculation provided that
a relatively strong attractive three-body potential, $S_\mathrm{3b} \approx -11.5$ MeV, is used.
This apparently uncharacteristic
large strength has no physical consequences, because the bulk of the
wave function is in the tail of this potential. This implies that a
large strength is needed to change the effective part of the
potential.

The other three-body structure investigated, the $\alpha$+$d$+$^2n$
configuration, describes reasonably well the experimentally known
spectrum of four states above the $\alpha$+$t$+$n$ threshold.  In
addition, the calculations predict a $2_2^+$ state and a $0_3^+$
state.  They could correspond to the two states that apparently are
experimentally known, but without assignments of angular momentum and
parity.  The most tempting interpretation is that the computed $2^+$
state is the lowest state at an excitation in $^8$Li of about
$7.1$~MeV, which is relatively close to the calculated $2_2^+$ energy,
and the $0^+_3$ state corresponds to the one observed experimentally
with an excitation energy of 9.0 MeV.

We can compare our results with available Monte Carlo calculations
\cite{Wiringa2000}, where the details deviate substantially from our results.
The number of levels are the same, but the order is different.  We
find one extra $0^+$ state above the three-body breakup threshold.
They get an extra $2^+$ state above the threshold $^7$Li*$+n$, but
below the breakup of $\alpha+t+n$.  They allow mixed cluster
structures, but all their states are dominated by the three-one
division, which must mean triton-neutron for the four valence
nucleons. The largest admixture of two-two cluster
(deuteron-dineutron) is in the highest $1^+$ and $3^+$ states, but
still less than $10\%$.

In total, we conclude that the $^8$Li states below $10$~MeV can be
understood as three-body structures, where the lowest in energy appear
as made of $\alpha$+$t$+$n$ and the higher states are build from
$\alpha$+$d$+$^2n$.  We can only compute widths for the resonances,
whereas most of the calculated states are bound, when confined to the
assumed three-body structure.  However, we can in detail describe the
structures, that is spin and parity, geometry and partial wave
decomposition of all states. The energies are relatively close to
experimental values, but must for accuracy be tuned with individual
three-body parameters.  The tuning leaves the structure essentially
unaffected, and, although it could be left out,  we find it rather satisfactory 
to see a full match with this final adjustment.  This is in contrast to shell model
or quantum Monte Carlo, where the energies of the states are
fluctuating around the measured values. The conclusion is that surprisingly 
small structureless adjustments of the calculations are sufficient to match 
the experimental information. We also predict the unknown quantum numbers 
for two excited states.

{\bf Acknowledgement:} This work has been partially supported by:
Grant PID2022-136992NB-I00 funded by MCIN/AEI/10.13039/501100011033.
We acknowledge continuous discussions and information from Karsten
Riisager, and for pointing us towards the $^8$Li spectrum in connection
with beta-delayed breakup of $^8$He.

\end{document}